\documentstyle[12pt]{article}
\normalbaselines 
\newcommand{\idty}{{\leavevmode{\rm 1\mkern -5.4mu I}}}
\newcommand{\Ir}{Z\!\!\!Z}
\newcommand{\Ibb}[1]{ {\rm I\ifmmode\mkern 
            -3.6mu\else\kern -.2em\fi#1}}
\newcommand{\ibb}[1]{\leavevmode\hbox{\kern.3em\vrule
     height 1.2ex depth -.3ex width .2pt\kern-.3em\rm#1}}
\newcommand{\Cx}{{\ibb C}}                                          
\newcommand{\Rl}{{\Ibb R}}                                          
\newcommand{\n}{n_i\otimes n_k}
\newcommand{\bi}{b_{ik}}
\newcommand{\A}{{\cal A}}
\newcommand{\U}{{\cal U}}
\newcommand{\C}{C_0\hat{\otimes}C_0}
\newcommand{\nn}{(n_i\otimes n_k-n_k\otimes n_i)}
\newcommand{\na}{n_i\wedge n_k}
\newcommand{\nik}{\nabla_{ik}}
\newcommand{\nki}{\nabla_{ki}}
\newcommand{\hten}{\hat{\otimes}}
\newcommand{\beq}{\begin{equation}}
\newcommand{\eeq}{\end{equation}}
\begin{document} 
\begin{center}
\vspace*{1.0cm}

{\LARGE{\bf Discrete Mathematics and Physics on the Planck-Scale
    exemplified by means of a Class of 'Cellular Network Models' and
    their Dynamics}} 

\vskip 1.5cm

{\large {\bf Manfred Requardt }} 

\vskip 0.5 cm 

Institut f\"ur Theoretische Physik \\ 
Universit\"at G\"ottingen \\ 
Bunsenstrasse 9 \\ 
37073 G\"ottingen \quad Germany

\end{center}

\vspace{1 cm}

\begin{abstract}
Starting from the hypothesis that both physics, in 
particular space-time and the physical vacuum, and the corresponding
mathematics are discrete on the Planck scale we develop a certain
framework in form of a class of '{\it cellular networks}' consisting of cells
(nodes) interacting with each other via bonds according to a certain
{\it 'local law'} which governs their evolution. Both the internal states of
the cells and the strength/orientation of the bonds are assumed to be dynamical
variables. We introduce a couple of candidates of such local laws
which, we think, are capable of catalyzing the unfolding of the
network towards increasing complexity and pattern formation.
In section 3 the basis is laid for a version of '{\it discrete
analysis}' on {\it 'graphs'} and {\it 'networks'}  which, starting
from different, perhaps more physically
oriented principles, manages to make contact with the much more
abstract machinery of Connes et al. and may complement the latter
approach. In section 4 several more advanced geometric/topological
concepts and tools are introduced which allow to study and classify
such irregular structures as (random)graphs and networks. We show in
particular that the systems under study carry in a natural way a 
{\it 'groupoid structure'}. In section 5 a, as far as we can see,
promising concept of
'{\it topological dimension}' (or rather: '{\it fractal dimension}')
in form of a '{\it degree of connectivity}' for graphs, networks and
the like is developed. It is then indicated how
this '{\it dimension}', which for continuous structures or regular
lattices being
embedded in a continuous background agrees with the ''usual''
notion of dimension (i.e. the respective embedding dimension) , may
vary dynamically as a result of a '{\it phase transition like}' change
of the '{\it connectivity}' in the network.   
\end{abstract} \newpage
\section{Introduction}
\noindent There exists a certain suspicion in parts of the scientific
community that nature may be "discrete" on the Planck scale. The
point of view held by the majority is however, at least as far as we
can see, that quantum theory as we know it holds sway more or less
unaltered down to arbitrarily small scales as an allembracing general 
principle, which is applied to a sequence of increasingly fine grained
effective field theories all the way down to, say, string field
theory. But even on that fundamental level one starts from strings moving in a
continuous background. It is then argued that "discreteness" enters
somehow through the backdoor via "quantisation".

The possibly most radical and heretical attempt, on the other
side, is it to try to generate both gravity and quantum theory as secondary and
derived concepts (in fact merely two aspects) of one and the same underlying 
more primordial theory instead of simply trying to quantise gravity,
which is the canonical point of view (see e.g. \cite{1}).

This strategy implies more or less directly that -- as gravity is
closely linked with the dynamics of (continuous) space-time -- the
hypothetical underlying more fundamental theory is supposed to live
on a substratum which does not support from the outset something like
continuous topological or geometrical structures. In our view these
continuous structures are expected to emerge as derived concepts via some sort
of coarse graining over a relatively large number of "discrete" more
elementary building blocks.

This program still leaves us with a lot of possibilities. For various
reasons, which may become more plausible in the course of the
investigation, we personally favor what we would like to call a
"cellular network" as a realisation of this substratum, the precise
definitions being given below. Without going into any details at the
moment some of our personal motivations are briefly the following:\\
i) These systems are in a natural way discrete, the local state space
at each site being usually finite or at least countable.\\
ii) Systems like these or their (probably better known) close
relatives, the "cellular automata", are known to be capable of
socalled "complex behavior", "pattern generation" and
"selforganisation" in general while the underlying dynamical laws are
frequently strikingly simple (a well-known example being e.g.
Conway's "game of life").\\
Remark: A beautiful introduction into this fascinating field is e.g.
\cite{2}. As a shorter review one may take the contribution of
Wolfram (l.c.). More recent material can be found in the proceedings
of the Santa Fee Institute, e.g. the article of Kauffman in \cite{3},
who investigates slightly different systems ("switching nets").\\
iii) Some people suspect (as also we do) that physics may be
reducible at its very bottom to some sort of "information processing
system" (cf. e.g. \cite{4,5}). Evidently cellular automata and the
like are optimally adapted to this purpose.\\
iv) In "ordinary" field theory phenomena evolving in space-time are
typically described by forming a fibre bundle over space-time (being
locally homeomorphic to a product). In our view a picture like this
can only be an approximate one. It conveys the (almost surely wrong) 
impression that
space-time is kind of an arena or stage being fundamentally different
from the various fields and phenomena which evolve and interact in
it. In our view these localised attributes, being encoded in the various
field values, should rather be attributes of the -- in the
conventional picture hidden -- infinitesimal neighborhoods of
space-time points, more properly speaking, neighborhoods in a medium
in which space-time is immersed as a lower dimensional "submanifold"
or, perhaps more properly expressed, coarse-grained ''super structure''.
To put it in a nutshell: We would prefer a medium in which what we
typically regard as irreducible space-time points have an internal
structure. To give a simple picture from an entirely different field:
take e.g. a classical gas, consider local pressure, temperature etc.
as collective coarse grained coordinates with respect to the
infinitesimal volume elements, regard then the microscopic degrees of
freedom of the particles in this small volume elements as the hidden
internal structure of the "points" given by the values of the above
collective coordinates (warning: this picture is of course not
completely correct as the correspondence between the values of
local pressure etc. and volume elements is usually not one-one).
It will turn out that a discrete structure as alluded to above is a
nice playground for modelling such features.\\[0.5cm]
Remark: There may exist certain resemblances between what we have said in
iv) and certain longstanding fundamental questions in pure mathematics
concerning the problem of the 'continuum', a catchword being e.g.
"non-standard analysis".\vspace{0.5cm}

A lot more could be said as to the general physical motivations and a
lot more literature could be mentioned as e.g. the work of
Finkelstein and many others (see e.g. \cite{6,7}.  For further references
cf. also the papers of Dimakis and M\"uller-Hoissen (\cite{8}) and the
deserving bibliography in \cite{Gibbs}.  Most
similar in spirit is in our view however the approach of 't Hooft
(\cite{9}).

In the following we will mainly concentrate on the developement of a
kind of discrete analysis on graphs and networks and the like and
compare it with other more abstract approaches. The unfolding phase
transition with its emergence of (proto) space-time together with the
necessary mathematical machinery will be described and analysed in a
companion paper (\cite{Planck}). In the latter paper one can find more
references, in particular concerning various branches of discrete
mathematics as e.g. '{\it random graphs}', '{\it discrete geometry}'
and advanced topics from '{\it combinatorics}', fields we expect to
play a major role in the future regarding the developement of an
appropriate framework as we have it in mind. A main achievement will
be the formulation of the concept of ''physical (proto) points''
within the framework of random graph theory. 

\section{The Concept of the "Cellular Network"}
While our primary interest is the analysis of various partly long
standing problems of current physics, which seem to beset physics
many orders away from the Planck regime, we nevertheless claim that
the understanding of the processes going on in the cellular network
at Planck level will provide us with strong clues concerning the
phenomena occurring in the "daylight" of
"middle-energy-quantum-physics". In fact, as Planck scale physics is
-- possibly for all times -- beyond the reach of experimental
confirmation, this sort of serious speculation has to be taken as a
substitution for experiments.

To mention some of these urgent problems of present day physics:\\
i) The unification of quantum theory and gravitation in general, and
in our more particular context: both the emergence of ''quantum
behavior'' and gravitation/space-time as two separate but related
aspects of the unfolding of the primordial network state, \\
ii) the origin of the universe, of space-time from "nothing" and
its very early period of existence,\\
iii) the mystery of the seeming vanishing of the 'cosmological
constant', which, in our view, is intimately related to the correct
understanding of the nature of vacuum fluctuations,\\
iv) the primordial nature of the "Higgs mechanism",\\
v) causality in quantum physics or, put differently, its strongly
translocal character,\\
vi) 'potential' versus 'actual' existence in the quantum world, the
ontological status of the wave function (e.g. of the universe))  and
the quantum mechanical measurement problem in general.

Most of these topics have been adressed in the thoughtful book of
S. Weinberg (\cite{weinberg}) and will be treated by us in much more
detail elsewhere in the near future. Therefore we refrain from making
more comments as to these fundamental questions at the moment apart
from the one remark that our approach will partly be based 
on the assumption that nature behaves or
can be imitated as a cellular network at its very bottom (\cite{10}).
It is however crucial for these investigations to have a sufficiently
highly developed form of ''discrete mathematics'' on graphs and
networks and the like at ones disposal. 
Therefore we will concentrate in the following mainly on establishing
the necessary (mostly mathematical) prerequisites on which the
subsequent physical investigations will be based.

This is the more so necessary because one of our central hypotheses
is that most of the hierarchical structure and fundamental building
blocks of modern physics come into being via a sequence of {\it unfolding
phase transitions} in this cellular medium. As far as we can see, the
study of phase transitions in cellular networks is not yet very far
developed, which is understandable given the extreme complexity of
the whole field. Therefore a good deal of work should be, to begin
with, devoted to an at least  qualitative understanding of this intricate
subject.

Furthermore discrete mathematics/physics of this kind is an
interesting topic as such irrespective of the applications mentioned
above which would justify a separate treatment of questions like the
following anyhow (cf. e.g. the very interesting paper by Mack,
\cite{mack}, where a complex of ideas is sketched to which we are
quite sympathetic) \\[0.5cm] 
{\bf 2.1 Definition(Cellular Automaton)}: A cellular automaton
consists typically of a fixed regular array of cells \{$C_i$\}
sitting on the nodes \{$n_i$\} of a regular lattice like, say, $\Ir^d$
for some d.  Each of the cells is characterized by its internal state
$s_i$ which can vary over a certain (typically finite) set $\cal S$
which is usually chosen to be the same for all lattice sites.

Evolution or dynamics take place in discrete steps $\tau$ and is
given by a certain specific 'local law' $ll$:
\begin{equation}s_i(t+\tau)=ll(\{s'_j(t)\},\;\underline{S}(t+\tau)=LL
(\underline{S}(t)) \end{equation}
where t denotes a certain "{\it clock time}" (not necessarily physical time), 
$\tau$ the elementary clock time interval, \{$s'_j$\} the
internal states of the nodes of a certain local neighborhood of the
cell $C_i$, {\it ll} a map:
\begin{equation}ll: {\cal S}^n \to \cal S 
\end{equation}
with n the number of neighbors occurring in (1), \underline{S}(t) the
global state at "time" t, {\it LL} the corresponding global map acting on
the total state space $X:=\{\underline{S}\}$. {\it LL} is called {\it
reversible} if it is a bijective map of X onto itself. \vspace{0.5cm}

Cellular automata of this type behave generically already very
complicated (see \cite{2}). But nevertheless we suspect they are
still not complicated enough in order to perform the specific type of
complex behavior we want them to do. For one, they are in our view
too regular and rigid for our purposes. For another, the occurring regular lattices
inherit quasi automatically such a physically important notion like
'{\it dimension}' from the underlying embedding space.

Our intuition is however exactly the other way round. We want to
generate something like dimension (among other topological notions)
via a dynamical process (of phase transition type) from a more
primordial underlying model which, at least initially, is lacking
such characteristic properties and features. 

There exist a couple of further, perhaps subjective, motivations
which will perhaps become more apparent in the following and which
result in the choice of the following primordial model
system:\\[0.5cm]
{\bf 2.2 Definition(Cellular Network)}: In the following we will
mainly deal with the class of systems defined below:\\
i) "Geometrically" they are {\it graphs}, i.e. they consist of nodes
\{$n_i$\} and bonds \{$\bi$\} where pictorially the bond $\bi$
connects the nodes $n_i$ and $n_k$ with $n_i\neq n_k$ implied (there
are graphs where this is not so), furthermore, to each pair of nodes
there exists at most one bond connecting them. In other words the
graph is 'simple' (schlicht). There is an intimate relationship
between the theory of graphs and the algebra of relations on sets. In
this latter context one would call a simple graph a set carrying a
homogeneous non-reflexive, (a)symmetric relation.

The graph is assumed to be {\it connected}, i.e. two arbitrary nodes
can be connected by a sequence of consecutive bonds, and {\it
regular}, that is it looks locally the same everywhere.
Mathematically this means that the number of bonds being incident
with a given node is the same over the graph ({\it 'degree' of a node}).
We call the nodes which can be reached from a given node by making
one step the {\it 1-order-neighborhood} ${\cal U}_1$ and by not more
than n steps ${\cal U}_n$.\\
ii) On the graph we implant a class of dynamics in the following way:\\[0.5cm]
{\bf 2.3 Definition(Dynamics)}: As for a cellular automaton each node
$n_i$ can be in a number of internal states $s_i\in \cal S$. Each
bond $\bi$ carries a corresponding bond state $J_{ik}\in\cal J$. Then
we assume:
\begin{equation} s_i(t+\tau)=ll_s(\{s'_k(t)\},\{J'_{kl}(t)\})
\end{equation}
\begin{equation} J_{ik}(t+\tau)=ll_J(\{s'_l(t)\},\{J'_{lm}(t)\})
\end{equation}
\begin{equation}
(\underline{S},\underline{J})(t+\tau)=LL((\underline{S},\underline{J})(
t))
\end{equation}
where $ll_s$, $ll_J$ are two mappings (being the same all over the
graph) from the state space of a local
neighborhood of a given fixed node or bond to $\cal S, J$, yielding the
updated values of $s_i$ and $J_{ik}$.\\[0.5cm]
Remarks: i) The theory of graphs is developed in e.g. \cite{11,12}. As
to the connections to the algebra of relations see also
\cite{schmidt}, for further references see \cite{Planck}. There are a
lot of concepts in graph theory which are
useful in our context, some of which will be introduced below where it
is
necessary. On the other hand we do not want to overburden this
introductory paper with to much technical machinery.  \\
ii) Synonyma for 'node' and 'bond' are e.g. 'site' and 'link' or
'vertex' and 'edge'.\\
iii) It may be possible under certain circumstances to replace or
rather emulate  a cellular network of the above kind by some sort of
extended cellular automaton (e.g. by replacing the bonds by
additional sites). The description will then however become quite
cumbersome and involved. \vspace{0.5cm}

What is the physical philosophy behind this picture? We assume the
primordial substratum from which the physical universe is expected to
emerge via a selforganisation process to be devoid of most of the
characteristics we are usually accustomed to attribute to something
like a manifold or a topological space. What we are prepared to admit
is some kind of "{\it pregeometry}" consisting in this model under
discussion of an irregular array of elementary grains and "direct
interactions" between them, more specifically, between the members of
the various local neighborhoods (see also \cite{Antonsen} for an
approach, which, while not following exactly the same lines, may be
similar in spirit).

It is an essential ingredient of our approach (in contrast to all the
others we are aware of) that the strength of
these direct interactions is of a dynamical nature and allowed to
vary. In particular it can happen that two nodes or a whole cluster
of nodes start to interact very strongly in the course of the
evolution and that this type of '{\it collective behavior}' persists
for a long time or forever (becomes '{\it locked in}') or, on the other
extreme, that the interaction between certain nodes becomes weak or even
vanishes.

It is not an easy task to select from the almost infinity of possible
models an appropriate subclass which we think has the potential of
displaying some or possibly all of the complex features (typically on
length scales far away from the Planck regime) we are confronted with
in ''ordinary'' (middle energy -- compared to the Planck scale -- )
quantum physics, and being, on the other side, sufficiently
transparent on, say, its natural primordial scale. We studied in fact
a lot of alternatives (we do not mention) and want to present in the
following some typical representatives of a certain class of models we
are presently favoring (for more details see \cite{Planck}).

Our guiding principles have been roughly the following: Most of the
cellular automaton rules being in use today (cf. e.g. \cite{2}) are of
a pronouncedly dissipative flavor. It is even frequently argued that
some kind of dissipation (or rather: shrinking of occupied 'phase space') is a
necessary prerequisite in order to have {\it 'attractors'} and, as a
consequence, pattern generation. We are not entirely convinced that the
arguments along these lines are really conclusive (for a class of
reversible automata see e.g. the book of Toffoli and Margolus,
\cite{toffoli})

In any case, as we want our model system to generate ''quantum
behavior'' on a, however, much coarser scale and if being in a certain
specific {\it 'phase'}, we consider it to be essential to implant a certain
propensity for {\it 'undulation'} in the class of local laws under
discussion. Furthermore, it turns out to be extremely useful - in
order to tame the horribly large quantum fluctuations occurring on
Planck scale, when probing into space-time regions of larger extension
- to incorporate a tendency to screen destructive fluctuations. These
''boundary conditions'' (among other considerations) led us to the
following types of model system, which are  only  simple representatives of
possibly a much larger class.

At each site $n_i$ there is sitting a one-dimensional discrete site
variable $s_i \in q\cdot\Ir$ with q, for the time being, a certain
elementary quantum. The bond variables $J_{ik}$ are, in the most simple
case, assumed to be two-valued, i.e. $J_{ik}\in \{\pm 1\}$.\\[0.5cm]
Remarks: i)For the time being we let the site variables range over the
full $\Ir$ in order not to complicate the already sufficiently
complicated reasoning further. It is of course possible to impose
certain boundary conditions (e.g. switching to a subgroup of $\Ir$) if
it turns out to be sensible (see the discussion in \cite{Planck}).\\
ii)In an extended model, which we will employ later on in order to
catalyze the {\it 'unfolding'} of the network together with the emergence of
space-time and gravitation, $J_{ik}$ can also take on the value $0$.\\
iii)In the next section on graphs we will give the graph an
{\it 'orientation'}, i.e. the bond $b_{ik}$ is assumed to point from $n_i$
to $n_k$ with $n_i$ initial node, $n_k$ terminal node, $b_{ki}$
denoting the same bond with reverse orientation (see Definition 3.1)
As a consequence, in order to be consistent, we assume:\\[0.5cm]
{\bf 2.4 Consequence}: 
\begin{equation}J_{ik}=-J_{ki}\end{equation}\vspace{0.5cm}

The physical idea behind this scheme is the following: If $J_{ik}$ is
positive an elementary quantum $q$ is transported in the elementary
'clock-time step' $\tau$ from node (cell) $n_i$ to $n_k$. Then the
first half of the local law reads:
\begin{equation}s_i(t+\tau)-s_i(t)=-q\cdot\sum_k J_{ik}\end{equation}
which is sort of a master or continuity equation.

What remains to be specified is the backreaction of the node states
onto the bond states. We make the following choice:
\begin{equation}\mbox{If}\quad s_i(t)>s_k(t)\quad \mbox{then}\quad
  J_{ik}(t+\tau)=+1\end{equation}
\begin{equation}\mbox{and hence}\quad
  J_{ki}(t+\tau)=-J_{ik}=-1\end{equation}
For the borderline case $s_i(t)=s_k(t)$ we have roughly two options
$B_1,B_2$ depending on the admissible state space of $J_{ik}$,
i.e. $\{-1,+1\}$ or $\{-1,0,+1\}$. In the former case we decree:
\begin{equation}J_{ik}(t+\tau)=J_{ik}(t)\quad\mbox{if}\quad
  s_i(t)=s_k(t)\end{equation}
and in the latter case:
\begin{equation}J_{ik}(t+\tau)=0\quad\mbox{if}\quad
  s_i(t)=s_k(t)\end{equation}

Introducing the signum function $sgn$ with
\begin{equation}sgn(x)=1,0,-1\quad \mbox{if}\quad x>0,=0,<0
\end{equation}
we then get:
$B_1)$ 
\begin{equation}J_{ik}(t+\tau)=-sgn(s_k(t)-s_i(t))+(1-|sgn(s_k(t)-s_i(t))|)J_{ik}(t)\end{equation}
$B_2)$
\begin{equation}J_{ik}(t+\tau)=-sgn(s_k(t)-s_i(t))\end{equation}

As already indicated above, in case we want to model the unfolding of
our ''network universe'' beginning with an extremely densly connected
initial state (a complete graph or simplex, say) with no genuine
physical neighborhood structure -- nodes are not experienced by each other
as near by or far away -- , we intensify the effect implemented in $B_2)$
and simulate what is called in catastrophy theory or in the realm of
self organisation a fold (in physics known as hystheresis):
C)
\begin{equation}i)\,J_{ik}(t+\tau)=-sgn(s_k(t)-s_i(t))\quad\mbox{if}\end{equation}
\begin{equation} |s_k(t)-s_i(t)|\ge\lambda_1\;\mbox{and}\;J_{ik}(t)\neq
  0\;\mbox{or}\;
  |s_k(t)-s_i(t)|>\lambda_2\;\mbox{and}\;J_{ik}(t)=0\end{equation}
with $\lambda_2>\lambda_1>0$ two critical parameters, indicating the
'{\it hysteresis interval}' $I_{\lambda}=[\lambda_1,\lambda_2]$
\begin{equation}ii)\;J_{ik}(t+\tau)=0\quad\mbox{if}\quad|s_k(t)-s_i(t)|<\lambda_1
\end{equation}
D) The same as in C) but with the roles of $\lambda_1,\,\lambda_2$
being interchanged, i.e. a bond is switched off if
$|s_k(t)-s_i(t)|>\lambda_2$ et cetera plus the above law for the
boundary case $s_k(t)=s_i(t)$: $J_{ik}(t+\tau)=J_{ik}(t)$.\\[0.5cm]
{\bf 2.5 Class of Local Laws}: For our purposes an admissible class of
local laws is given by the representatives $A)$ plus $B_1,B_2$, $C$ or
$D$.\\[0.5cm]
Remarks:i)The reason why we do not choose the ''current'' $q\cdot
J_{ik}$ proportional to the ''voltage difference'' $(s_i-s_k)$ as
e.g. in Ohms's law is that we favor a non-linear(!) network which is capable of
self-excitation and self-organisation rather than self-regulation
around a relatively uninteresting equilibrium state! The balance
between dissipation and amplification of spontaneous fluctuations has
however to be carefully chosen (''complexity at the edge of chaos'')\\
ii)We presently have emulated these local network laws on a
computer. As far as we can see, the most promising law is variant
D). In any case, it is fascinating to observe the enormous capability
of such intelligent networks to find attractors very rapidly, given
the enormous accessible phase space (for more details see
\cite{Planck}).\\
iii)In the class of laws discussed so far a direct
bond-bond-interaction is not yet implemented. We are prepared to
incorporate such a contribution if it turns out to be necessary. In
any case it is not entirely obvious how to do it in a sensible way,
stated differently, the class of possible physically sensible interactions is
perhaps not so numerous.\\ 
iv)Note that -- in contrast to e.g. euclidean lattice field theory
-- the socalled '{\it clock time}' $t$ is, for
the time being, not standing on
the same footing as, say, potential ''coordinates'' in the network (e.g.
curves of nodes/bonds). We suppose anyhow that socalled '{\it physical
time}' will emerge as sort of a secondary collective variable in the
network, i.e. being different from the clock time (while being of
course functionally related to it).

In our view this is consistent with the spirit of relativity. What
Einstein was really teaching us is that there is a (dynamical)
interdependence between what we experience as space respectively time, not
that they are absolutely identical! In any case the assumption of an
overall clock time is at the moment only made just for convenience in
order to make the model system not too complicated. If our
understanding of the complex behavior of the network dynamics
increases, this assumption may be weakened in favor of a possibly
local or/and dynamical clock frequency. A similar attitude should be
adopted concerning concepts like '{\it Lorentz-(In)Covariance}' which
we also consider as '{\it emergent}' properties (needless to say that
it is of tantamount importance to understand the way how these
patterns do emerge from the relatively chaotic background which will
be attempted in future work).
\vspace{0.5cm}

As can be seen from the definition of the cellular network it
separates quite naturally into two parts of a different mathematical
and physical nature. The first one comprises part i) of definition
2.2, the second one part ii) and definition 2.3. The first one is
more static and "geometric" in character, the latter one conveys a
more dynamical and topological flavor as we shall see in the
following. On the other side, it turns out to be useful to consider
graphs with the bonds itself being switched on or off. this will be
done in \cite{Planck}.

We begin in section 3 with a representation of what may be called
discrete analysis on graphs and networks. This is followed in section
4 by making the first steps into an investigation of certain possible
dynamical processes in networks of
the defined type which have the character of  phase transitions or collective
behavior and may induce {\it
dimensional change}. Most importantly we develop a physically
appropriate concept of '{\it dimension}' for such irregular discrete
structures which may be of importance in a wider context.

\section{Discrete Analysis on Networks}
At first glance one would surmise that as an effect of discreteness
something like a network will lack sufficient structure for such a
discipline to exist, but this is not so. Quite the contrary, there
are intimate and subtle relations to various recent branches of pure
mathematics as e.g. '{\it cyclic (co)homology}', '{\it noncommutative
de Rham complexes}', '{\it noncommutative geometry}' in general and
the like (see e.g. \cite{13}-\cite{16}, as a beautiful and concise
survey we recommend also \cite{coque}).

It is the general aim of these recent developements to generate
something like a geometrical and differentiable structure within 
certain mathematical contexts which traditionally are not considered
to support such structures.  Particularly simple examples are
discrete sets of, say, points, e.g. lattices. In a series of papers
Dimakis and M\"uller-Hoissen have applied the general abstract
machinery to models like these which have a possible bearing to, say,
lattice field theory etc. (see e.g. \cite{8} and further references
there).

The fundamental object in these approaches is typically the socalled
'{\it universal differential algebra}' or '{\it differential
envelope}' which can be canonically constructed over any associative
algebra and which is considered to be a generalisation or surrogate
(depending on the point of view) of a differential structure in the
ordinary cases.

As the adjective 'universal' may already indicate, this scheme
, paying tribute to its
universality and generality, is sometimes relatively far away from
the concrete physical models one is perhaps having in mind. In the
case of networks, for example, the inevitable starting point of this
approach would be the
'{\it maximally connected}' network or graph (also called a '{\it
complete graph}' or in algebraic topology a '{\it simplex}'), i.e. any
two nodes are directly connected by a bond.

As a consequence, the construction is lacking, at least initially,
something which is of tantamount importance in physical models, i.e.
a natural and physically motivated neighborhood structure (much more
about this concept can be found in e.g. \cite{Planck}). Typically
the interesting physical models are relatively lowly connected, which
implies that they usually exhibit a pronounced feeling of what is near
by or far away on the network.

One can of course pull this general structure down to the level of
the models one may have in mind by imposing '{\it relations}' between
various classes of '{\it differential forms}' employing a general
result that each differential calculus over an algebra is isomorphic
to the universal one modulo a certain 'differential ideal', but anyway, given a
concrete model this approach is relatively abstract and perhaps not
the most transparent and direct one. While being mathematically
correct we want nevertheless to make some reservations as to its
concrete meaning (or rather: interpretation) as far as specific models
are concerned
(e.g . in particular for networks and graphs); our main criticism
being that it may introduce a host of ''unnatural'' and artificial
relations among the constituents of the model which have no foundation in
the physical structure of the model, see the end of this section. Furthermore,
 it stresses more
the global algebraic relations and perhaps not so much the inherent
topological/geometrical content of the given model theory. 

Stated differently, networks and graphs behave only ''mildly
non-commutative'' or rather 'non-local'. On the other hand they convey
a lot more of extra structure (as most models do), which is not
automatically implemented in the general algebraic scheme but has to
be brought to light by scrutinizing the specific model class under discussion.

To put it in a nutshell: one can either go the way ''top down'',
starting from some branch of non-commutative geometry and realize in
the course of time that e.g. discrete sets or graphs may serve as
certain model systems for this abstract algebraic scheme, or one may
start from some concrete physical speculations and ideas about the 
supposed fine
structure of the physical vacuum and space-time as a dynamical unfolding
network and then make ones way ''bottom up'' observing that part of
the emerging mathematical structure  may be viewed as a variant of
non-commutative geometry.

We will follow the latter route in this paper and think the 
two philosophies may
complement each other even if, coming from different directions, one
may sometimes end up at formally closely related concepts.

We begin with the introduction of some useful concepts borrowed from
algebraic topology and also known from graph theory (as to this we
recommend the beautiful book of Lefschetz, \cite{17}).

In a first step we have to give the graph an '{\it orientation}':\\[0.5cm]
{\bf 3.1 Definition(Orientation)}: With the notions defined in
definition 2.2 we say the bond $\bi$ points from node $n_i$ to node
$n_k$, the bond $b_{ki}$ from $n_k$ to $n_i$. We call $n_i$, $n_k$
initial and terminal node of $\bi$ respectively. We assume the up to
now formal relation:
\begin{equation} \bi=-b_{ki} \end{equation}
Remark: Note that orientation in the above (mathematical) sense is
different from what is understood in many applications as '{\it
directed bond}' in a network (as e.g. in typical "Kauffman nets",
\cite{3}). There a directed bond can typically "transport", say, a
message only in one given fixed direction. That is, nets of this type
behave, in physical terms, pronouncedly anisotropic locally. The
definition 3.1, on the other side, is rather implementing something
like the orientation of curves.\\[0.5cm]
{\bf 3.2 Definition(Chain Complexes)}: We introduce, to begin with,
the two vector spaces $C_0$, $C_1$ whose elements, {\it zero- and
one-chains} are defined by up to now formal sums
\begin{equation} \underline{c}_0 :=\sum f_in_i\quad \underline{c}_1
:=\sum g_{ik}b_{ik} \end{equation} \\[0.5cm]
where the $f_i$'s and $g_{ik}$'s range over a certain given field or
ring, of in the simplest cases numbers (i.e.$\Ir$,$\Rl$,$\Cx$), the
$n_i$'s and $b_{ik}$'s serve as generators of a free module.\\[0.5cm]
Remarks: i) Evidently one could in a further step choose much more general
spaces as candidates from which the $f_i's$ or $g_{ik}$ are to be
taken like, say, groups or manifolds.\\
ii) Furthermore, for the time being, the $f_i$'s and $g_{ik}$'s
should not be confused with the $s_i$'s and $J_{ik}$'s introduced in
section 2. The $f_i$'s and $g_{ik}$'s are e.g. allowed to vanish
outside a certain given cluster of nodes in various calculations or,
put differently, it may be convenient to deal only with certain
subgraphs.\\
iii) The spaces $C_0$, $C_1$ are in fact only the first two members
of a whole sequence of spaces.\\[0.5cm]
{\bf 3.3 Definition (Boundary)}: we now define a {\it boundary
operator} by
\begin{equation} \delta \bi:=n_k-n_i \end{equation}
which by linearity induces a linear map from $C_1$ to $C_0$:
\begin{equation} \delta: C_1\ni \sum g_{ik}\bi\to \sum
g_{ik}(n_k-n_i)\in C_0 \end{equation}
The kernel, $Z_1$ of this map, the 1-chains without '{\it boundary}',
consist of the '{\it 1-cycles}'. A typical example is a '{\it loop}',
i.e. a sequence of bonds, $\sum_{\nu}b_{i_{\nu}k_{\nu}}$ s.t.
$k_{\nu}=i_{\nu+1}$ and $k_n=i_1$. (However, not every cycle is a
loop!).\\[0.5cm]
{\bf 3.4 Definition(Coboundary)}: Analogously we can define the coboundary
operator as a map from $C_0$ to $C_1$:
\begin{equation} dn_i:=\sum_k b_{ki} \end{equation}
where the sum extends over all bonds having $n_i$ as terminal node,
and by linearity:
\begin{equation} d: \sum_i f_in_i \to \sum_i f_i\,(\sum_k b_{ki})
\end{equation}
Remarks:i)In algebraic topology 'cotheory' is frequently defined on
'dual spaces'. At the moment we do not make this distinction.\\
ii)To avoid possible formal complications, we always assume the
'degree' of the nodes to be uniformly bounded away from
infinity. These matters could however be more appropriately dealt with
after the introduction of suitable metrics, norms and related
topological concepts.\vspace{0.5cm}

We will now show that these two operations, well known in algebraic
topology, can be fruitfully employed to create something like a
discrete calculus. Evidently, the 0-chains can as well be considered as
functions over the set of nodes; in this case we abbreviate them by
f,g etc. (if necessary, chosen from  a certain subclass of 0-chains ${\cal A} \subset  
C_0$, e.g. of '{\it finite support}', $L^1$, $L^2\ldots$). $\cal A$ is
trivially a module over itself (pointwise multiplication) freely
generated by the nodes $\{n_i\}$ which can be identified with the
'{\it elementary functions}' $e_i:=1\cdot n_i$.

With $b_{ik}=-b_{ki}$ we can write the action of $d$ on $f$
differently, thus making  its slightly hidden
meaning more transparent:\\
With $b_{ki}=1/2(b_{ki}-b_{ik})$ we get
\begin{equation}\sum_i(f_i \sum_k b_{ki})=1/2\sum_{ik}
  (f_k-f_i)\,b_{ik}\end{equation}
i.e:\\[0.5cm]
{\bf 3.5 Observation}: \begin{equation} df=d(\sum_i
f_in_i)=1/2\cdot\sum_{ik}(f_k-f_i)\,\bi \end{equation} 
\beq d(\idty)=d(\sum n_i)=\sum_{ik} b_{ik}=1/2\sum_{ik}(1-1)b_{ik}=0\eeq
\vspace{0.5cm}

We have still to show to what extent the operation d defined above
has the properties we are expecting from an (exterior) derivation.
The really crucial property in the continuum case is the (graded)
Leibniz rule. This is in fact a subtle and interesting point. To see
this we make a short aside about how discrete differentiation is
usually expected to work.

Take the following definition:\\[0.5cm]
{\bf 3.6 Definition (Partial Forward Derivative and Partial
Differential at Node (i))}:
\begin{equation} \nabla_{ik}f(i):=f(k)-f(i) \end{equation}
where $n_i,n_k$ are '{\it nearest-neighbor-nodes}', i.e. being
connected by a bond $\bi$.\\[0.5cm]
{\bf 3.7 Observation}: 
\begin{eqnarray}
\nabla_{ik}(f\cdot g)(i) & = & (f\cdot g)(k)-(f\cdot g)(i)
\nonumber\\
                         & = & \nabla_{ik}f(i)\cdot
g(i)+f(k)\cdot\nabla_{ik}g(i)\\
                         & = &
\nabla_{ik}f(i)g(i)+f(i)\nabla_{ik}g(i)+\nabla_{ik}f(i)\nabla_{ik}g(i)
\end{eqnarray}
In other words the "derivation" $\nabla$ does {\bf not} obey the ordinary(!)
Leibniz rule. In fact, application of $\nabla$ to, say, higher powers
of f becomes increasingly cumbersome (nevertheless there is a certain
systematic in it). One gets for example (with $q:=\nabla_{ik}$):
\begin{equation}q(f_1\cdots f_n)=\sum_i f_1\cdots q(f_i)\cdots
  f_n+\sum_{ij} f_1\cdots q(f_i)\cdots q(f_j)\cdots f_n+\ldots+
  q(f_1)\cdots q(f_n) \end{equation} 

Due to the discreteness of the formalism and, as a consequence, the
inevitable bilocality of the derivative there is no chance to get
something as a true Leibniz rule on this level. (That this is
impossible has also been stressed arguing from a different point of view in
e.g. example 2.1.1 of \cite{14}).\\[0.5cm]
Remark: We will come back to the non-Leibnizean character of $\nabla$
below when establishing a {\it duality} between $d$ and $\nabla$. It
is in fact a rather interesting relation even from a purely algebraic
point of view, as it is a structural relation known in algebraic
topology as {\it 'Cuntz algebra'} (cf. \cite{coque} or \cite{Cuntz};
see also the following section). Before however
doing that we will further clarify the role of $d$.\vspace{0.5cm} 

In some sense it is considered to be one of the merits of the
abstract algebraic framework (mentioned at the beginning of this
section) that a graded Leibniz rule holds in that generalized case
almost automatically.
The concrete network model under investigation offers a good
opportunity to test the practical usefulness of concepts like these.

To write down something like a Leibniz rule an important structural
element is still missing, i.e. the multiplication of node functions
from, say, some $\cal A$ with the members of $C_1$, in other words a
'{\it module structure}' over $\cal A$. One could try to make the
following definition:
\begin{equation} f\cdot\bi:=f(i)\cdot\bi\quad \bi\cdot
f:=f(k)\cdot\bi \end{equation}
and extend this by linearity.

Unfortunately this "definition" does not respect the relation
$\bi=-b_{ki}$. We have in fact:
\begin{equation} f(i)\bi=f\cdot\bi=-f\cdot b_{ki}=-f(k)b_{ki}=f(k)\bi
\end{equation}
which is wrong in general for non-constant $f$!

Evidently the problem arises from our geometrical intuition which
results in the natural condition $\bi=-b_{ki}$, a relation we however
want to stick to. On the other side we can extend or embed our
formalism algebraically in a way which looses the immediate contact with
geometrical evidence but grants us with some additional mathematical
structure. This is in fact common mathematical practice and a way to
visualize e.g. the {\it 'universal differential algebra'} in {\it
  'non-commutatine geometry'} (see e.g. \cite{coque}). We want however
to complement this more algebraic extension scheme by a, as we think,
more geometric one below. 

We can define another relation between nodes, calling two nodes
related if they are connected by a bond with a fixed built-in direction 
from the one to the other (i.e. not an orientation as above!). We
express this in form of a {\it tensor product} structure. In the
general tensor product $C_0\otimes C_0$ we
consider only the subspace $\C$ spanned by the elements $\n$ with
$n_i,n_k$ connected by a bond (i.e. $i\neq k$!) and consider $\n$ to be unrelated to
$n_k\otimes n_i$, i.e. they are considered to be linearly independent
basis elements.\\[0.5cm]
{\bf 3.8 Observation}: There exists an isomorphic embedding of $C_1$ onto
the subspace generated by the antisymmetric elements in $\C$, i.e:
\begin{equation} \bi\to \nn=:\na\quad (\mbox{with}\; i\neq k) \end{equation}
generate an isomorphism by linearity between $C_1$ and the
corresponding subspace $C_0\wedge C_0 \subset \C$.\\[0.5cm]
Proof: Both $\bi$ and $\na$ are linearly independent in there 
respective vector spaces apart from the relation
$b_{ik}=-b_{ki}$,$n_i\wedge n_k=-n_k\wedge n_i$. \vspace{0.5cm}

In contrast to $C_1$ the larger $\C$ now supports a non-trivial and natural
bimodule structure:\\[0.5cm]
{\bf 3.9 Observation/Definition (Bimodule)}: We can now define
\begin{eqnarray} f\cdot (\n) & := & f(i)(\n)\\
          (\n)\cdot f        & := & f(k)(\n)
\end{eqnarray}
and extend this by linearity to the whole $\C$, making it into a
bimodule over some ${\cal A} \subseteq C_0$.\\[0.5cm]
Remarks:i) Equivalently one could replace $\n$ by $e_i\otimes e_k$, the
corresponding elementary functions. If one now identifies
$e_i\otimes e_k$ with the abstract symbols $e_{ik}:=e_ide_k$ employed
in \cite{8}, one may establish a link to perhaps more abstract but
related approaches (see also the end of this section).\\
ii) Another case in point is our definition $dn_i=\sum_k b_{ki}$ and
the representation $da=\idty\otimes a-a\otimes\idty$ which is employed
within the context of the universal differential algebra. With
$\idty=\sum_i 1\cdot n_i$ and Observation 3.8 the close relation
becomes immediately apparent, i.e:
\begin{equation}dn_i=\sum_k b_{ki}=\sum_k (n_k\otimes n_i-n_i\otimes  n_k)=
(\sum_k n_k)\otimes n_i-n_i\otimes(\sum_k n_k)\end{equation}
and with $n_i\to e_i$, $\sum e_i=\idty$ this equals
\begin{equation}dn_i=\idty\otimes n_i-n_i\otimes\idty\end{equation}
The complete equivalence between $n_i\otimes n_k$ or $e_i\otimes e_k$
and $e_{ik}:=e_ide_k$ can then be seen with the help of Observation
3.9 and the above representation for $dn_i$ or $de_i$.
\\[0.5cm]
{\bf 3.10 Lemma}: As a module over $\cal A$,\, $\C$ is generated by
$C_0\wedge C_0$.\\[0.5cm]
Proof: It suffices to show that every $\n$ can be generated this way.
\begin{equation} n_i\cdot\nn=\n \end{equation}
as $n_i\cdot n_k=0$ for $i\neq k$.\\[0.5cm]
Remark: Note that this is not so in general, i.e. with
$da=\idty\otimes a-a\otimes\idty$ one gets only 
$bda=b\otimes a-ba\otimes\idty$. In our particular context the $n_i$'s
are however a basis for the algebra ${\cal A}$.
\vspace{0.5cm}

With the $\bi$ so embedded in a larger space and identified with
\begin{equation}
\nn=\na 
\end{equation}
we are in the position to derive a graded Leibniz
rule on the module (algebra) $\A$. Due to linearity and the structure
of the respective spaces it suffices to show this for products of
elementary functions $e_i=n_i$. The same relation could of course be
directly verified in a slightly more elegant way by regrouping 
\begin{equation}d(f\cdot g)=\sum_i f_ig_idn_i=-\sum_i f_ig_i(\sum_k \nn)
\end{equation}
appropriately, employing {\bf 3.9 Observation}.
We in fact have:\\[0.5cm]
($i\neq k$ not nearest neighbors):\begin{equation}d(n_i\cdot
n_k)=0,\,dn_i\cdot n_k=n_i\cdot dn_k=0 \end{equation}
($i\neq k$ nearest neighbors):\begin{equation}d(n_i\cdot
n_k)=d(0)=0\;\;\mbox{and} \end{equation}
\begin{eqnarray}dn_i\cdot n_k+n_i\cdot dn_k & = &
-(\sum_{k'}b_{ik'})\cdot n_k-n_i\cdot(\sum_{i'}b_{ki'})\\ & = &
-\bi\cdot n_k-n_i\cdot b_{ki}\\
& = &- \{\nn n_k+n_i(n_k\otimes n_i-n_i\otimes
n_k)\}\\ & = &-\{n_i\otimes n_k-n_i\otimes n_k\}=0 \end{eqnarray} 
$(i=k)$:
\begin{equation} d(n_i^2)=d(n_i)=-\sum_k\bi\;\;\mbox{and} \end{equation}
\begin{eqnarray} dn_i\cdot n_i+n_i\cdot dn_i & = &
-(\sum_k\bi)\,n_i-n_i(\sum_k\bi)\\ & = & -\sum_k\nn=-\sum_k\bi=dn_i
\end{eqnarray}
{\bf 3.11 Conclusion}: As a map from the bimodule ${\cal A} \subseteq
C_0$ to the  bimodule $\C$ generated by the elements $\bi$
over $\A$ the map d fulfills the Leibniz rule, i.e:
\begin{equation} d(f\cdot g)=df\cdot g+f\cdot dg \end{equation}

From the above we see also that functions, i.e. elements from $\A$
and bonds or differentials of functions do no longer commute (more
specifically, the two possible ways of imposing a module structure
could be considered this way). We have for example:\\[0.5cm]
{\bf 3.12 Commutation Relations}:\\
($i\neq k$ not nearest neighbors): \begin{equation} n_i\cdot
dn_k=dn_k\cdot n_i=0 \end{equation}
($i\neq k$ nearest neighbors). \begin{eqnarray} n_i\cdot
dn_k+dn_k\cdot n_i & = & -\sum_{i'}\{n_i(n_k\otimes
n_{i'}-n_{i'}\otimes n_k)\\ & + & (n_k\otimes
n_{i'}-n_{i'}\otimes n_k)\,n_i\}\\ & = & \nn=\bi \end{eqnarray}
($i=k$): \begin{equation} n_i\cdot dn_i+dn_i\cdot n_i=-\sum_k\bi=dn_i
\end{equation}

Another important relation we want to mention is the followng:
$\delta\,d\,f$ is a map from $C_0\to C_0$ and reads in detail:\\[0.5cm]
{\bf 3.13 Observation (Laplacian)}: \begin{equation}
\delta\,d\,f=-\sum_i(\sum_k f(k)-n\cdot f(i))\,n_i=:-\Delta\,f \end{equation}
with $n$ the number of nearest neighbors of $n_i$ and $\sum_k$ extending
over the nearest neighbors of $n_i$ (both being node dependent in
general!)
\\[0.5cm]  
Proof: \begin{eqnarray} \delta\,d\,f & = &
1/2\sum_{ik}(f(k)-f(i))(n_k-n_i)\\ & = &
1/2\sum_{ik}(f(k)\,n_k+f(i)\,n_i-f(i)\,n_k-f(k)\,n_i)\\ & = &
-\sum_i(\sum_kf(k)-n\cdot f(i))\,n_i \end{eqnarray}

Before we will introduce additional geometric concepts in the next
section, which will carry the flavor of our specific model class
(i.e. graphs and networks), we want to conclude this section by
addressing briefly the case of a complete graph in order to exhibit
the close resemblance of our approach in this particular case with the
general abstract construction.

With a simplex as underlying space we do not have to worry about
forming arbitrary ''products''. The universal differential algebra
$\Omega({\cal A})$ over an associative unital algebra ${\cal A}$ (with
$d\idty:=0$) is a $\Ir$-graded algebra generated by $a_i,\,da_i$. Its
'{\it words}' can be normalized to $a_0da_1\cdots da_n$ with the help
of the Leibniz rule. Products of such monomials are then defined by 
concatenation and can be put into normal form by repeated application
of the Leibniz rule:
\begin{equation}(a_0da_1\cdots da_n)\cdot(b_0db_1\cdots db_m)=
a_0da_1\cdots(da_n\cdot b_0)\cdot db_1\cdots db_m\end{equation}
and with
\begin{equation}da_n\cdot b_0=d(a_n\cdot b_0)-a_n\cdot
  db_0\end{equation}
the product of $\Omega({\cal A})$ restricted to $\Omega^0({\cal
  A}):={\cal A}$ being the ordinary product.\\[0.5cm]
{\bf 3.14 Observation}:i) In our particular context (complete network
or complete graph) with the $n_i$,$dn_i$ as building blocks one can
easily show that e.g.
\beq n_{i_1}\cdot dn_{i_2}\cdots dn_{i_k}=(n_{i_1}\cdot
dn_{i_2})\cdot(n_{i_2}\cdot dn_{i_3})\cdots (n_{i_{k-1}}dn_{i_k})\eeq
holds for $i_j\neq i_{j+1}$ (this is a consequence of the Leibniz rule
and $n_i\cdot n_j=\delta_{ij}\cdot n_i$) and should be compared with
the approach presented in \cite{8}.\\
ii) On the other side, an expression like $(n_1dn_2)\cdot (n_idn_k)$
and expressions containing such a term with $i\neq 2$ are zero, as it
is equal to
\beq -(n_1\cdot n_2)dn_idn_k+n_1d(n_2\cdot n_i)dn_k\eeq
with $n_1\cdot n_2=n_2\cdot n_i=0$ in $\A$ by
assumption.\vspace{0.5cm}

With our ''tensor-product realisation'' we have ($i\neq k$):
\beq n_idn_k=\n\eeq
It remains to define the realisation of abstract concatenation within
this representation.\\[0.5cm]
Remark: Note that there do exist (to some extent) structurally
different realisations of the abstract universal differential algebra
(in particular concerning the implementation of the product rule), see
e.g. section 2 in \cite{coque}.\\[0.5cm]
Observation 3.14, which shows that a ''standard basis element'' of
$\Omega^n$ (i.e. $n$ differentials) corresponds to $n$ products
\beq (n_{i_1}\otimes n_{i_2})\cdot(n_{i_2}\otimes n_{i_3})\cdots
(n_{i_{n-1}}\otimes n_{i_n})\eeq
suggests the following rule:
\beq (n_i\otimes n_j)\cdot(n_k\otimes n_l):=n_i\otimes (n_j\cdot
n_k)\otimes n_l\eeq
yielding\\[0.5cm]
{\bf 3.15 Corollary}:\beq n_1dn_2\cdots dn_k=n_1\otimes
n_2\otimes\cdots\otimes n_k\eeq
On the other hand, the above shows also that our algebra contains a
lot of '{\it zero divisors}'.\\[0.5cm]
Remark: This is one of the reasons why we will argue in the following
section to regard the model under discussion rather as a natural
candidate for a '{\it groupoid}'.\vspace{0.5cm}

As a last remark one should perhaps say some words about the
completeness of the above basis elements. We showed in Lemma 3.10 that
in our particular case the submodule $\Omega^1$, generated by the
$dn_i$'s over ${\cal A}$, is already the full tensor product
$\A\hat{\otimes}\A$ spanned by the elements $n_i\otimes
n_k$ with $i\neq k$! As a consequence we have\\[0.5cm]
{\bf 3.16 Corollary}: 
\beq
\Omega^k:=\Omega^1\otimes_{\A}\cdots\otimes_{\A}\Omega^1=\{a_0da_1\ldots
da_k\}\eeq
(where the $da_i$ are not necessarily distinct) equals the tensor
product
\beq \A\hat{\otimes}\cdots\hat{\otimes}\A\;
\mbox{$(k+1)$-times}\eeq
spanned by $\{n_{i_0}\otimes\cdots\otimes n_{i_k}\}\; i_{\nu}\neq
i_{\nu+1}$.
\\[0.5cm]
Proof: The only thing which remains to be shown is that expressions
like e.g. $da\cdot da$ or $dn_1\cdot dn_1$ can be spanned by 
$n_{i_0}\otimes n_{i_1}\otimes n_{i_2}\;i_{\nu}\neq i_{\nu+1}$. We
have
\beq dn_1=(\idty\otimes n_1-n_1\otimes\idty)=(\sum_{k\neq 1}n_k\otimes
n_1-n_1\otimes\sum_{k\neq 1}n_k)\eeq
In the product $dn_1\cdot dn_1$ a lot of terms vanish, the
non-vanishing ones yielding:
\beq dn_1\cdot dn_1=-\sum_{k\neq 1}n_1\otimes n_k\otimes
n_1-\sum_{k\neq 1}\sum_{k'\neq 1}n_k\otimes n_1\otimes n_{k'}\eeq
which has the desired structure.\vspace{0.5cm}

With the help of the concrete realisation within our graph model one
is able to give these expressions an interpretation by means of purely
geometrical (graph) properties which will to some extent be done in
the following section.

\section{Some more advanced Geometrical Concepts}
Having now established the first steps in setting up this particular
version of discrete calculus one could proceed in various directions.
First, one can develop a discrete Lagrangian variational calculus,
derive Euler-Lagrange-equations and Noetherian theorems and the like
and compare this approach with other existing schemes in discrete
mathematics.  

Second, one can continue the above line of reasoning and proceed to
more sophisticated geometrical concepts and set them into relation to
existing work of mostly a more abstract flavor. For the time being we
would like to follow this latter route and briefly sketch in a couple
of subsections various, as we think, interesting aspects of our model system.

\subsection{(Co)Tangential Spaces, Cuntz Algebra, Connections etc.}
The philosophy underlying non-commutative geometry is that
e.g. individual points of, say, a manifold have to be dispensed with
and replaced by some equivalent of the algebra of functions over the
manifold. On the other side networks are, as was already mentioned
above, only mildly non-commutative and carry still a pronounced local
structure (even the notion of points make still sense),
which it may be advisable to implement in the discrete calculus.

As a consequence we will develop, as in the case of ordinary
manifolds, differentials and partial derivatives in
parallel as dual concepts. This may be, in our opinion, perhaps one of
the differences of our approach as compared to other existing (more
abstract) work in the field.

A characteristic feature of network calculus is its
non-locality. According to Definition 3.6 the partial derivatives
$\nik$ act locally(!), i.e. one can consider them as acting at node
$n_i$. On the other side, this is not so for the $\bi$; it leads to
inconsistencies if one tries to relate them somehow to a definite
node. We are however free to introduce the dual concept with respect
to the $\nik$'s and define: \\[0.5cm]
{\bf 4.1 Definition ((Co)Tangential Space)}:\\i) We call the space
spanned by the $\nabla_{ik}$ at node $n_i$ the tangential space
$T_i$.\\ ii) Correspondingly we introduce  the space spanned by the
$d_{ik}$ at node $n_i$ and call it
the cotangential space $T_i^{\ast}$ with the $d_{ik}$ acting as linear
forms over $T_i$ via:
\begin{equation} <d_{ik}|\nabla_{ij}>=\delta_{kj} \end{equation}\\[0.5cm]
{\bf 4.2 Definition/Observation}: Higher tensor products of differential forms
at a node $n_i$ can now be defined as {\it multilinear forms}:
\begin{equation} <d_{ik_1}\otimes\cdots\otimes
d_{ik_n}|(\nabla_{il_1},\cdots,\nabla_{il_n})>:=\delta_{k_1l_1}
\times\cdots{\times} \delta_{k_nl_n} \end{equation}
and linear extension.\vspace{0.5cm}

In a next step we extend these concepts to functions $f\in
C_0$ and {\it 'differential operators'} or {\it 'vector fields'} $\sum a_{ik}\nik$
and make the following interpretation:\\[0.5cm]
{\bf 4.3 Interpretation}: Instead of the above identification of
$b_{ik}$ with $\nn$ we may equally well identify $\bi$ with
$d_{ik}-d_{ki}$.\vspace{0.5cm}

We have to check whether this is a natural(!) identification.\\[0.5cm]
{\bf 4.4 Observation}: Vector fields $v:=\sum a_{ik}\nik$ are assumed
to act on functions $f=\sum f_in_i$ in the following manner:
\begin{equation}v(f):=\sum a_{ik}(f_k-f_i)n_i\end{equation}
i.e. they map $C_0\to C_0$.\\[0.5cm]
{\bf 4.5 Corollary}: Note that this implies:
\begin{equation}\nik n_k=n_i\quad \nik n_i=-n_i\end{equation}
\begin{equation}\nki n_k=-n_k\quad \nki n_i=n_k\end{equation}\\[0.5cm]
{\bf 4.6 Observation}: {\it 'Differential forms'} $\omega =\sum
g_{ik}d_{ik}$ act on vector fields $v=\sum a_{ik}\nik$ according to:
\begin{equation}<\omega|v>=\sum
  g_{ik}a_{ik}n_i\end{equation}\vspace{0.5cm}

With these definitions we can calculate $<df|v>$ with
\begin{equation}df=1/2\sum (f_k-f_i)\bi=\sum
  (f_k-f_i)d_{ik}\end{equation}
according to Interpretation 4.3. Hence:
\begin{equation}<df|v>=<\sum (f_k-f_i)d_{ik}|\sum a_{ik}\nik>=\sum
  (f_k-f_i)a_{ik}n_i\end{equation}
which equals:
\begin{equation}(\sum a_{ik}\nik)(\sum
  f_in_i)=v(f)\end{equation}\\[0.5cm]
{\bf 4.7 Consequence}: Our geometric interpretation of the algebraic
objects reproduces the relation:\\
i) \begin{equation}<df|v>=v(f)\end{equation}
known to hold in ordinary differential geometry, as is the case for
the following relations, and shows that the definitions made above
seem to be natural.\\
Furthermore vector and covector fields are left modules under the
action of ${\cal A} \subseteq C_0$:\\
ii) \begin{equation}(\sum f_in_i)(\sum a_{ik}\nik):=\sum f_ia_{ik}\nik
\end{equation}
iii)\begin{equation}(\sum f_in_i)(\sum g_{ik}d_{ik}):=\sum
f_ig_{ik}d_{ik}\end{equation}
iv) As was the case with the $\n$ the $d_{ik}$ generate also in a
natural way a right module over ${\cal A}$.\vspace{0.5cm}

We mentioned above that the partial derivatives $\nik$ do not
obey the ordinary (graded) Leibniz rule in contrast to the operator
$d$. In the latter case this is however only effected by embedding the
''natural'' geometric objects in a bigger slightly more abstract
space. The structural relation the $\nik$'s are actually obeying (see 
Definition 3.7) is known from what is called by algebraic topologists a
{\it 'Cuntz algebra'} (cf. \cite{coque} and \cite{Cuntz}); which
however does occur there in
a different context. With $q:=\nik$ we have:\\[0.5cm]
{\bf 4.8 Observation (Cuntz algebra)}:
\begin{equation}q(f\cdot g)=q(f)\cdot g+f\cdot q(g)+q(f)\cdot
  q(g)\end{equation}
and analogously for vector fields $\sum a_{ik}\nik$.\\
With $u:=1+q$ we furthermore get:
\begin{equation}u(f\cdot g)=u(f)\cdot u(g) \end{equation}
and
\begin{equation}q(f\cdot g)=q(f)\cdot g+u(f)\cdot q(g)\end{equation}
i.e. a {\it 'twisted derivation'} with $u$ an endomorphism from $\A$
to $\A$.\vspace{0.5cm}

As is the case with $\nik$, the product rule for higher products can be inferred
inductively:
\begin{equation}q(f_1\cdots f_n)=\sum_i f_1\cdots q(f_i)\cdots
  f_n+\sum_{ij} f_1\cdots q(f_i)\cdots q(f_j)\cdots f_n+\ldots+
  q(f_1)\cdots q(f_n) \end{equation}\\[0.5cm]
{\bf 4.9 Conclusion}: The above shows that, in contrast to classical
  differential geometry, we have a dual pairing between vector and
  covector fields with the vector fields acting as twisted(!)
  derivations on the node functions while the corresponding
  differential forms obey the graded Leibniz rule like their classical
  counterparts.\vspace{0.5cm}

Another important geometrical concept is the notion of '{\it
  connection}' or '{\it covariant derivative}'. Starting from the
  abstract concept of (linear) connection in the sense of Koszul it is
  relatively straightforward to extend this concept to the
  non-commutative situation, given a '{\it finite  projective module}'
  over some algebra ${\cal A}$ (instead of the sections of a vector
  bundle over some manifold ${\cal M}$, the role of $\A$ being played
  by the functions over ${\cal M}$; see e.g. \cite{13} or
  \cite{coque}, as to various refinements and improvements cf. e.g. 
\cite{Dubois} and further references given there).

Without going into any details we want to briefly sketch how the
concept of connection can be immediately implemented in our particular
model without referring to the more abstract work. We regard (in a
first step) a connection as a (linear) map from the fields of tangent
vectors to the tensor product of tangent vectors and dual differential
forms as defined above (and having certain properties).\\[0.5cm]
{\bf 4.10 Definition/Observation(Connection)}: A field of connections,
$\Gamma$, is defined at each node $n_i$ by a (linear) map:
\beq \nik\to\gamma_{kl}^j(n_i)\cdot \nabla_{ij}\otimes d_i^l\eeq
where the index $i$ plays rather the role of the ''coordinate'' $n_i$,
the index $l$ is raised in order to comply with the summation
convention. The $\gamma_{kl}^j$'s are called '{\it connection
  coefficients}'. The corresponding '{\it covariant derivative}'
$\nabla$ obeys the relations:\\
i)\beq \nabla(v+w)=\nabla(v)+\nabla(w)\eeq
ii)\beq \nabla(f\cdot v)=v\otimes df+\nabla(v)\cdot f\eeq
iii)\beq \nabla(\nik)=\Gamma(\nik)\quad
df=\sum(f_k-f_i)d_{ik}\eeq\\[0.5cm]
Remark: The tensor product in ii) is understood as the pointwise
product of fields at each node $n_i$, i.e. $\nabla_{ik}$ going with
$d_{ik}$. This is to be contrasted with the abstract notion of tensor
product in e.g. the above differential algebra $\Omega(\A)$ which does
not(!) act locally, the space consisting of, say, elements of the kind
$n_1\otimes n_2\otimes\cdots\otimes n_k$. These diferrent parallel
structures over the same model shall be scrutinized in more detail
elsewhere. Note that the above extra locality structure is a
particular property of our model class and does not (openly) exist in
the general approach employing arbitrary '{\it projective modules}'
respectively {\it differential algebras}.

\subsection{The Groupoid Structure of the (Reduced) Algebraic Differential
  Calculus on  Graphs and its Geometric Interpretation}

In this subsection we will provide arguments that the algebraic
differential calculus, more specifically: the kind of calculus
introduced in section 3, should most naturally be regarded as an
example of a '{\it groupoid}', in particular if the underlying graph
is not a simplex, in other words: if the differential algebra is not
the universal one (this is to be contrasted with the treatment of
'{\it reduced calculi}' by M\"uller-Hoissen et al; see the papers
mentioned above).

In contrast to the universal differential algebra (associated with a
simplex) where every two
nodes are connected by a bond, this is not so for the
'{\it reduced}' calculus over a non-complete graph. As a
consequence certain algebraic operations are
straightforward to define in the former approach. However, descending
afterwards  to the lower-connected perhaps more realistic models is tedious
in general and not always particularly transparent. That is, this
method does not always really save calculational efforts (for a
discussion of certain simple examples see \cite{8}). It even may lead
to (in our view) ''unnatural''(!) results as we want to show below.

By the way, the mathematical "triviality " of the differential envelope is
reflected by the trivialty of the corresponding '{\it (co)homology
groups}' of the maximally connected graph (simplex). This trivialty
is then broken by deleting graphs in the reduction process.

To mention a typical situation: Take e.g. the subgraphs of a graph G
consisting of, say, four nodes $n_i,n_k,n_l,n_m$
and all the bonds between them which occur in G (i.e. a '{\it section
  graph}' or '{\it full subgraph}'). In the case G being a simplex 
(i.e. non-reduced case) all these
subgraphs are geometrically/topologically equivalent. An important
consequence of this is that the four nodes can be connected by a
{\it 'path'}, i.e. a sequence of consecutive bonds, each being passed
only once, the effect being that one can naturally reconstruct the
subsimplices in the corrresponding algebraic scheme via '{\it
  concatenation}'
of the four nodes in arbitrary order, e.g:
\begin{equation}n_1\otimes n_2\otimes n_3\otimes n_4  \end{equation}
In the same way all the
higher subsimplices of order, say, $n$ can be reconstructed via
concatenation of the $n$ nodes (none occurring more than once). The
reason why it is sufficient to concatenate (geometrically) only at
the extreme left and right of a '{\it word}' stems exactly from the
simplex-character of each section graph!

In typical reduced cases, however, all this is no longer the case; the
combinatorial topology becomes non-trivial. To give an example: Take
as G a graph containing a $4$-node section graph  with bonds existing only
between, say, $n_1\,n_4$, $n_2\,n_4$, $n_3\,n_4$. I.e., one has the
1-forms $b_{14},\;b_{24},\;b_{34}$ or:
\begin{equation} n_1\otimes n_4,\;n_2\otimes n_4,\;n_3\otimes n_4
\end{equation}
but there is no obvious way to generate or reconstruct the
corresponding section graph by concatenating the above pieces
sequentially(!) without passing certain bonds twice, that is, the only
way to represent the section graph algebraically via concatenation is
by means of $n_1\hten n_4\hten n_2\hten n_4\hten n_3$.\\[0.5cm]
Remark: This shows that a natural correspondence between algebraic
concepts and geometric ones is perhaps not so immediate as long as 
multiplication is simply defined by concatenation at the end of
words. It is one of our aims in this context to generalize
algebraic multiplication in a natural (more geometrical) way so that arbitrary
subgraphs can be multiplied and composed more freely (i.e. not simply
'{\it sequentially}') while keeping as much algebraic structure as
possible. This shall however be done elsewhere. To prepare the ground
we want to adress in the following two subsections\\
i) the natural '{\it groupoid structure}' of our algebraic
construction\\
ii) the problem of unavoidable {\it ''unnatural'' relations} if one
tries to represent the reduced calculus as a quotient of the universal
differential algebra.\vspace{0.5cm}

We have shown above (see Corollary 3.16) that the elements of a
differential algebra, $a_0da_1\ldots da_k$, can be naturally
represented within our graph context as elements of a restricted
tensor product, i.e. with building blocks,
$n_{i_0}\otimes\cdots\otimes n_{i_k}\in \Omega^k=C_0\hten\cdots\hten
C_0\;\mbox{((k+1)-times)}$ where the hat means $n_{i_{\nu}}\neq
n_{i_{\nu+1}}$. In the same sense multiplication can be defined as the
concatenation of such sequences.\\[0.5cm]
{\bf 4.11 Observation}: The algebraic structure of the abstract
differential algebra can be represented ''geometrically'' as the
concatenation of admissible '{\it bond sequences}' described by the
string of nodes $n_{i_0}\ldots n_{i_k}$ with the proviso that
consecutive nodes are connected by a bond.\\[0.5cm]
Remarks: i) In the following it is always understood that $\otimes$
occurs only between nodes which are connected by a bond.\\
ii) The above shows that the ring which can be formed in this way has
typically a lot of {\it 'zero divisors'}, that is, admissible strings
being concatenated so that the end node of the first is different
from the initial node of the second string.\\
iii) Note however that an object like $n_0\otimes\cdots\otimes
n_k\in \Omega_k$ should not be confused with $n_0\otimes
n_1+n_1\otimes n_2+\cdots n_{k-1}\otimes n_k\in \Omega_1$. Both
elements could be associated geometrically with a bond sequence or
path leading from $n_0$ to $n_k$, but they are of an entirely 
different algebraic (and
geometric(!)) character. In the former case the pieces of the bond
sequence are concatenated multiplicatively in the latter case they are
composed additively.\vspace{0.5cm}

Remark ii) shows that the above algebraic structure, based on the
multiplication of arbitrary strings of nodes (leading to a ring or
algebra structure) is perhaps not(!) the most natural one. In the
following we would like to suggest a, at least in our view, more
natural structure.\\[0.5cm]
{\bf 4.12 Definition (Groupoid)}: A '{\it groupoid}' $\Gamma(G,B)$
consists of two sets, $G,B$ and two maps, $r,s$ from $G$ to $B$ ('{\it
  range}' and {\it source}') with $B$ kind of a '{\it base
  space}'. The elements of $G$ obey the following law of
composition. With $g_1,g_2\in G$:
\beq (g_1,g_2)\to g_1\cdot g_2\in G\eeq
provided that $r(g_1)=s(g_2)$.\\
The multiplication fulfills the following properties:\\
i) it is associative if either of the products
\beq (g_1\cdot g_2)\cdot g_3,\quad g_1\cdot(g_2\cdot g_3)\eeq
is defined\\
ii) each $g$ has a left and right identity $e^L_g,\;e^R_g$ so that
\beq g\cdot e^R_g=e^L_g\cdot g=g\eeq
iii) each $g$ has a left and right inverse $g^{-1}$ with
\beq g\cdot g^{-1}=e^L_g,\quad g^{-1}\cdot g=e^R_g\eeq\vspace{0.5cm}

We would like to note that a slightly different (but basically
equivalent) characterisation is in
use (see e.g. \cite{13},chapter II.5):\\[0.5cm]
{\bf 4.13 Definition (Groupoid, second version)}: $B$ is now
considered as a distinguished subset of $G$. $s,r$ fulfill the
properties:\\
i)\beq s(g_1\cdot g_2)=s(g_1)\,,\,r(g_1\cdot g_2)=r(g_2)\eeq
ii)\beq s(x)=r(x)=x\quad\mbox{for}\quad x\in B\subset G\eeq
iii)\beq s(g)\cdot g=g\,,\,g\cdot r(g)=g\eeq
iv) each $g$ has a two-sided inverse $g^{-1}$ with
\beq g\cdot g^{-1}=s(g)\;,\;g^{-1}\cdot g=r(g)\eeq 
while the associative law is assumed to hold as in Definition
4.12.\\[0.5cm]
Remarks:i) In many respects and many (modern) applications the notion
of groupoid appears to be more natural than the group-concept.\\
ii) Some recent literature can be found in \cite{groupoid}, note in
particular the groupoid home page of the University of
Colorado.\vspace{0.5cm}

We concluded section 3 with the observation that in the algebraic
approach to our network calculus the essential building blocks consist
of strings of bonds or admissible sequences of nodes,
$n_0\otimes\cdots\otimes n_k$, with each pair of consecutive nodes
connected by a bond. Multiplication of such strings is defined if the
end node of the first string coincides with the initial node of the
second string, their product being
\beq n_0\otimes\cdots\otimes n_k\otimes\cdots\otimes
n_{k+m}\;\mbox{with}\;n_{k+j}=n'_j\eeq

It is now fairly evident that, after suitable modifications, these
strings fulfill the axioms of a groupoid. The identifications are as
follows:\\[0.5cm]
{\bf 4.14 Representation of Groupoid Axioms}:\\
i)\beq G=\mbox{Set of strings},\,\{n_0\otimes\cdots\otimes n_k\}\eeq
ii)\beq B=\mbox{Set of nodes},\,\{n_i\}\eeq
iii) source map $s$, range map $r$:
\beq s(n_0\cdots n_k)=n_0,\;r(n_0\cdots n_k)=n_k\eeq
iv) the multiplication of strings, $g_1\cdot g_2$, is defined if
$r(g_1)=s(g_2)$\\
v) the multiplication is associative and fulfills all the axioms of
Definition 4.13\\
vi) One can identify the nodes with corresponding ''zero-strings'' and
define:
\beq s(n_i)=n_i=r(n_i)\eeq
\beq n_0\cdot(n_0\otimes\cdots\otimes n_k)=(n_0\otimes\cdots\otimes
n_k)\cdot n_k=n_0\otimes\cdots\otimes n_k\eeq
hence each $g$ has a left and right identy (which are, however,
different in general!)\\
vii) with
\beq g^{-1}=(n_0\cdots n_k)^{-1}:=n_k\cdots n_0\eeq
(i.e. the string oriented in the opposite way) we define 
\beq (n_0\cdots n_k)\cdot(n_k\cdots n_0)=n_0\otimes\cdots
n_{k-1}\otimes n_k\otimes n_{k-1}\otimes\cdots\otimes
n_0:=n_0=s(g)\eeq
\beq (n_k\cdots n_0)\cdot(n_0\cdots n_k):=n_k=r(g)\eeq
Remark: vii) together with the associative law implies that a
substring of the form $g\cdot g^{-1}$ which occurs in a string of the
form $\alpha\cdot g\cdot g^{-1}\cdot\beta$ can be replaced by
$s(g)$. This is very natural from a geometric point of view as it
represents a path which is traversed consecutively in both possible
directions.

\subsection{The Problem of ''Unnatural'' Relations in the purely
  Algebraic Approach}

In this last subsection we want to adress the problem of ''unnatural
relations'' in the general algebraic approach which, starting from the
general universal differential algebra (in our context the complete
graph or simplex), exploits the fact that each (algebraic)
differential calculus can be got via dividing the universal one  by a
socalled '{\it differential ideal}'. In the following we will argue
that this general statement needs some
qualifications as physical model theories do usually convey a lot more
structure than can be inferred from the abstract uninterpreted
axiomatic scheme.

To begin with, we want to illustrate with the help of an example how
one would naturally proceed in the construction of ''higher
dimensional'' geometric objects within the specific context we have
developed above without resorting to the abstract algebraic
machinery. 

These objects may be considered as equivalents
of the building blocks of piecewise affine or triangulated smooth
manifolds. Following our general philosophy of creating
geometric/topological concepts for {\it 'non-standard'} spaces (cf. in
particular the next section), one can try to catch the abstract
essence of a notion like surface or volume in developing the following
scheme which we however only sketch, for the time being, with the help
of an example.

The map $d$ mapped a node $n_i$ onto the sum over bonds $\sum b_{ki}$
with endpoint $n_i$, where the oriented $b_{ki}$ was the
antisymmetric combination $d_{ki}-d_{ik}$ or $n_{ki}-n_{ik}$. In the
same sense one can proceed geometrically if the graph has an
appropriate structure:\\[0.5cm]
{\bf 4.15 Definition}: A triangle in a graph is a triple of nodes
$n_1,n_2,n_3$ with $n_1\,n_2,n_2\,n_3,n_3\,n_1$ connected by bonds.\\
In case $n_1,n_2,n_3$ form a triangle it should be
algebraically/geometrically realized (as is the case with the bond
$b_{ik}$) as a geometric object being a totally antisymmetric
combination of elements of $\Omega_2$ in the following sense:
\begin{equation}s_{123}:=\sum_{per}
  \sigma(per)n_{i_1i_2i_3}\end{equation}
with $\sigma(per)$ the signum of the corresponding permutation of
$\{1,2,3\}$. \\[0.5cm]
{\bf 4.16 Observation}: With this definition the triangle has exactly
two possible orientations , i.e:
\begin{equation}s_{123}=-s_{321}\end{equation}
which corresponds with the two possible orientations of the path
around the triangle.\\[0.5cm]
Remarks: i) It is of course possible that $n_{12},n_{23}$ and therefore
$n_{123}$ can be built while $n_1\,n_3$ are not connected s.t. they do
not form a triangle. A fortiori a graph need not have triangles.\\
ii) Following similar lines one can build also higher geometric
objects.\vspace{0.5cm}

{\bf 4.17 Extension of d}: The geometric idea behind the attempt to
extend in a next step the range of the map $d$ is that it should
relate a bond (edge) with an appropriate combination of the triangles
being incident with it, i.e:
\begin{equation}d(b_{12})=\sum_i s_{12i}\end{equation}

This will cause certain algebraic problems of a general nature for
arbitrary, in particular non-regular, graphs (see the above remark) as
typical relations known to hold e.g. in {\it 'simplicial cohomology'}
like $d\circ d=0$ will not hold automatically. The necessary
preconditions could however be analyzed more systematically but we
shall not make the
attempt to do this here in full detail apart from providing an
argument concerning the failure of the relation $d\circ d=0$ (see
below). 
Instead of that we will address another topic which is in fact
closely related to these problems and will make contact with a
different approach which starts from the universal differential
envelope, i.e., given the class of nodes, from the, in our language,
complete graph (simplex); see e.g. \cite{8}. 

In that extremely regular situation matters are rather smooth and can
directly be taken over from the general case of the universal
differential algebra $\Omega(\A)$ defined over an arbitrary algebra
$\A$. In our notation one defines e.g:
\begin{equation}d_u(n_{1\ldots
    k})=\sum_i(\sum_{\nu=0}^k (-1)^{\nu}n_{1\ldots i\ldots
    k})\end{equation}
with $\nu$ denoting the place of the insertion of node $n_i$,
beginning with $\nu=0$, i.e. before $n_1$ and, as always, consecutive
nodes being different understood. $\sum_i$ runs over the nodes which
    are linked with both $n_{\nu}$ and $n_{\nu+1}$. One shows immediately that
e.g. $d_u\circ d_u=0$ on $\Omega$ as there occurs always the same
    string twice in the sum with an even respectively an odd power of
    $(-1)$. \\[0.5cm]
Remarks:i) Note that each term on the rhs is well defined since all
the nodes are connected with each other.\\
ii) With $d_u$ we denote the {\it universal} derivation on the
    complete graph.\vspace{0.5cm}

We remarked at the beginning of this section that it is a general
result that each differential calculus (more specifically: a certain
differential algebra) over a given algebra $\A$ is isomorphic to the
universal one divided by a certain {\it 'differential ideal'}. This
was exploited in \cite{8} to construct a differential calculus on
certain simple examples of {\it 'reduced'} (smaller) differential
algebras.

In our specific context of networks and graphs we may translate this
general result in the following way: Let $\Omega^u=\sum \Omega^u_k$ be
the universal differential algebra with $\Omega^u_k$ consisting of the
$(k+1)$-fold tensor products of arbitrary $(k+1)$-tuples of nodes. We
can define a projector $\Pi$ which projects $\Omega^u$ onto
$\Omega=\sum \Omega_k$ with $\Omega_k$ consisting of the $(k+1)$-fold
{\it admissible} tensor products (bond sequences) of connected nodes
in our actual network (i.e. the reduced graph). We have:
\begin{equation}\Pi(n_0\otimes\cdots\otimes n_k)=0\end{equation}
if $(n_0,\ldots ,n_k)$ is not admissible
\begin{equation}\Pi(n_0\otimes\cdots\otimes
  n_k)=n_0\otimes\cdots\otimes n_k\end{equation}
if $(n_0,\ldots ,n_k)$ is admissible.\\[0.5cm]
{\bf 4.18 Consequence}: We have 
\begin{equation}\Pi=\Pi^2\; , \;
  \Omega^u=\Pi\Omega^u+(\idty-\Pi)\Omega^u\end{equation}
with $\Pi\Omega^u=\Omega$.\\
We can define
\begin{equation}d:=\Pi\circ d_u\circ\Pi\end{equation}
which leaves $\Omega$ invariant but in general $d\circ d\neq 0$ in
contrast to $d_u\circ d_u=0$.\vspace{0.5cm}

The reason is the following: We can make $Ker(\Pi)$ into a two-sided
ideal $I$ consisting of the elements $n_{0\ldots k}$ having at least
one pair of consecutive nodes {\it not} being connected by a
bond in the reduced graph. This ideal $I$ is however {\it not}
invariant under the action of $d_u$!
A closer analysis shows that $d_u(n_{0\ldots k})\notin I$ if $d_u$ creates
{\it 'insertions'} between non-connected neighbors in the reduced graph
s.t. non-admissible elements become admissible,
i.e. connected.

The non-vanishing of $d\circ d$ can be understood with the help of the
following argument:\\
In the reduced graph the following can happen. Apply $d$ to a given
string which yields e.g. an insertion between node $n_i$ and node
$n_{i+1}$, e.g. \quad$\ldots n_in_{\nu}n_{i+1}\ldots$\quad with a weight
$(-1)^i$. Applying $d$ again may yield another admissible insertion of
the type \quad$\ldots n_in_{\mu}n_{\nu}n_{i+1}\ldots$\quad coming with the weight
$(-1)^i\cdot (-1)^i$ provided that $n_{\mu}$ is connected with $n_i$
and $n_{\nu}$.\\
On the other hand the ''counterterm'' with the weight
$(-1)^i\cdot(-1)^{i+1}$ may be missing as it can happen that $n_{\mu}$
is connected with $n_i$ and $n_{\nu}$ but not(!) with $n_i$ and
$n_{i+1}$ so that $\ldots n_in_{\mu}n_{i+1}\ldots$ does not show up in
the first step in contrast to the analogous term with $n_{\nu}$.
\\[0.5cm]
{\bf 4.19 Observation}: In general there exist elements $n_{0\ldots
  k}\in Ker(\Pi)$ s.t. $\Pi(d_u(n_{0\ldots k}))\neq 0$, in other
words, $I$ is in general not left invariant by $d_u$.\\This is the
reason for e.g. the non-vanishing of $d\circ d$.\\
If one wants to make $\Omega$ a real differential algebra one has to
enlarge $I$!\\[0.5cm]
{\bf 4.20 Consequence}: The ideal $I'=I+d_u\circ I$ is invariant under
$d_u$ and $d$ defines a differential algebra on the smaller algebra
$\Omega^u/I'\subset \Omega$ with $\Omega=\Omega^u/Ker(\Pi)$.\\[0.5cm]
(That $I'$ is an ideal left invariant by $d_u$ is easy to prove with
the help of the property $d_u\circ d_u=0$).\vspace{0.5cm}

So much so good, but in our view there exists a certain
problem. $\Omega^u/I'$ is the algebra one automatically arrives at if
one defines the homomorphism $\Phi$ from $\Omega^u$ to the reduced
differential algebra in the following canonical way:
\begin{equation}\Phi:\; n_i\to n_i\quad d_un_i\to dn_i\end{equation}
i.e. under the premise that $d$ defines already another differential
algebra. It is in this sense the general result mentioned above has to
be understood.

On the other hand this may lead to a host of, at least in our view,
unnatural relations in concrete examples as e.g. our network which
may already carry a certain physically motivated interpretation going
beyond being a mere example of an abstract differential algebra. Note
e.g. that in our algebra $\Omega$ an element like $n_{123}$ is
admissible (i.e. non-zero) if $n_1,n_2$ and $n_2,n_3$ are
connected. $n_{123}$ may however arise from a differentiation process
(i.e. from an insertion) like $d_u(n_{13})$ with $n_1,n_3$ not(!)
connected.

This is exactly the situation discussed above:
\begin{equation}n_{13}\in I \quad\mbox{but}\quad d_u(n_{13})\notin
  I\end{equation}
Dividing now by $I'$ maps $d_u(n_{13})$ onto zero whereas there may be
little\\ physical/geometric reason for $n_{123}$ or a certain
combination of such admissible elements being zero in our
network.\\[0.5cm]
{\bf 4.21 Conclusion}: Given a concrete physical network $\Omega$ one
has basically two choices. Either one makes it into a full-fledged
differential algebra by imposing further relations which may however
be unnatural from a physical point of view and very cumbersome for
complicated networks. This was the strategy
e.g. followed in \cite{8}.\\
Or one considers $\Omega$ as the fundamental object and each
admissible element in it being non-zero. As a consequence the
corresponding algebraic/differential structure on $\Omega$ may be less
smooth at first glance ($dd\neq 0$ in general), but on the other side
more natural.\\
At the moment we refrain from making a general judgement whereas we
would probably prefer the latter choice. 

\section{Intrinsic Dimension in Networks, Graphs and other Discrete Systems}
There exist a variety of concepts in modern mathematics which
generalize the notion of '{\it dimension}' one is accustomed to in
e.g. differential topology or linear algebra. In fact, '{\it
topological dimension}' and related concepts are notions which seem to
be even closer to
the underlying intuition (cf. e.g. \cite{18}).

Apart from the purely mathematical concept there is also a physical
aspect of something like dimension which has e.g. pronounced effects
on the behavior of, say, many-body-systems, especially their
microscopic dynamics and, most notably, their possible '{\it phase
transitions}'.

But even in the case of e.g. lattice systems they are usually
considered as embedded in an underlying continuous background space
(typically euclidean) which supplies the concept of ordinary
dimension so that the {\it 'intrinsic dimension'} of the discrete array itself
does usually not openly enter the considerations.

Anyway, it is worthwhile even in this relatively transparent
situation to have a closer look on where attributes of something
like dimension really come into the physical play. Properties of
models of, say, statistical mechanics  are almost solely derived from
the structure of the microscopic interactions of their constituents.
This is more or less the only place where dimensional aspects enter
the calculations.

Naive reasoning might suggest that it is the number of nearest
neighbors (in e.g. lattice systems) which reflects in an obvious way
the dimension of the underlying space and influences via that way the
dynamics of the system. However, this surmise, as we will show in the
following, does not reflect the crucial point which is considerably
more subtle.

This holds the more so for systems which cannot be considered as being
embedded in a smooth regular background and hence do not get their
dimension from the embedding space. A case in point is our primordial
network in which Planck-scale-physics is assumed to take
place. In our approach it is in fact exactly the other way round:
Smooth space-time is assumed to emerge via a {\it phase transition} or a
certain {\it cooperative behavior} and
after some '{\it coarse graining}' from this more fundamental
structure.\\[0.5cm]
{\bf 5.1 Problem}: Formulate an intrinsic notion of dimension for
model theories without making recourse to the dimension of some
continuous embedding space.\vspace{0.5cm}

In a first step we will show that graphs and networks as introduced
in the preceding sections have a natural metric structure. We have
already introduced a certain neighborhood structure in a graph with
the help of the minimal number of consecutive bonds connecting two
given nodes.

In a connected graph any two nodes can be connected by a sequence of
bonds. Without loss of generality one can restrict oneself to '{\it
paths}'. One can then define the length of a path (or sequence of
bonds) by the number l of consecutive bonds making up the
path.\\[0.5cm] 
{\bf 5.2 Observation/Definition}: Among the paths connecting two
arbitrary nodes there exists at least one with minimal length which
we denote by $d(n_i,n_k)$. This d has the properties of a '{\it
metric}', i.e:
\begin{eqnarray} d(n_i,n_i) & = & 0\\ d(n_i,n_k) & = &
d(n_k,n_i)  \\d(n_i,n_l) & \leq & d(n_i,n_k)+d(n_k,n_l) \end{eqnarray}
(The proof is more or less evident).\\[0.5cm]
{\bf 5.3 Corollary}: With the help of the metric one gets a natural
neighborhood structure around any given node, where ${\cal U}_m(n_i)$
comprises all the nodes, $n_k$, with $d(n_i,n_k)\leq m$, 
$\partial{\cal U}_m(n_i)$
the nodes with $d(n_i,n_k)=m$. \vspace{0.5cm}

This natural neighborhood structure enables us now to develop
the concept of an intrinsic dimension on graphs and networks. To this
end one has at first to realize what property really matters
physically (e.g. dynamically) independently of the model or embedding
space. \\[0.5cm]
{\bf 5.4 Observation}: The crucial and characteristic property of,
say, a graph or network which may be associated with something like
dimension is the number of '{\it new nodes}' in ${\cal U}_{m+1}$ compared
to ${\cal U}_m$ for m sufficiently large or $m\to \infty$. The deeper
meaning of this quantity is that it measures the kind of '{\it
wiring}' or '{\it connectivity}' in the network and is therefore a
'{\it topological invariant}'.\\[0.5cm]
Remark: In the light of what we have learned in the preceding section
it is tempting to relate the number of bonds branching off a node,
i.e. the number of nearest neighbors or order of a node, to something
like dimension. On the other side there exist quite a few different
lattices with a variety of number of nearest neighbors  in, say, two- or three-
dimensional euclidean space. What however really matters in physics
is the embedding dimension of the lattice (e.g. with respect to phase
transitions) and only to a much lesser extent the number of nearest
neighbors. In contrast to the latter property dimension reflects the degree of
connectivity and type of wiring in the network. \vspace{0.5cm}

In many cases one expects the number of nodes in ${\cal U}_m$ to grow like
some power D of m for increasing m. By the same token one expects the
number of new nodes after an additional step to increase proportional
to $m^{D-1}$. With $|\,\cdot\,|$ denoting number of nodes
we hence have:
\begin{equation} |{\cal U}_{m+1}|-|{\cal U}_m|=|\partial {\cal U}_{m+1}|=f(m)
\end{equation}
with
\begin{equation} f(m)\sim m^{D-1} \end{equation}
for m large.\\[0.5cm]
{\bf 5.5 Definition}: The intrinsic dimension D of a homogeneous 
(infinite) graph is given by 
\begin{equation} D-1:=\lim_{m\to \infty}(\ln f(m)/\ln m)\; \mbox{or}
\end{equation}
\begin{equation} D:=\lim_{m\to \infty}(\ln |{\cal U}_m|/\ln m)
\end{equation}
provided that a unique limit exists!\\
What does exist in any case is $\liminf$ respectively $\limsup$ which
can then be considered as upper and lower dimension. If they coincide
we are in the former situation. By '{\it homogeneous}' we mean that 
$D$ does not depend on the reference point  \\[0.5cm]
Remarks:i) One might expect that '{\it regularity}', i.e. constant
node degree, plus certain other conditions imply homogeneity but this is
a highly non-trivial question. There are e.g. simple examples of
regular graphs which do not ''look the same'' around every node, that
is, regularity alone is not sufficient    \\
ii) Furthermore other definitions of dimension are possible,
e.g. incorporating the bonds instead of the nodes. These various
possibilities and their mutual interdependence are presently under
study, the details being published elsewhere.\\
iii) For practical purposes one may also introduce a notion of local
dimension around certain nodes or within certain regions of a not
necessarily regular graph if the above limit is approached
sufficiently fast.\\
iv) This becomes particularly relevant in cases where we treat
networks in a more random fashion (e.g. '{\it random graphs}',
cf. \cite{Planck}).\vspace{0.5cm}

That this definition is reasonable can be seen by applying it to
ordinary cases like regular translation invariant lattices. It is
however not evident that such a definition makes sense for arbitrary
graphs, in other words, a (unique) limit point may not always
exist. It would be tempting to characterize the conditions which
entail that such a limit exists. We, however, plan to do this elsewhere.
\\[0.5cm]
{\bf 5.6 Observation} For regular lattices D coincides with the
dimension of the euclidean embedding space $D_E$.\\[0.5cm]
Proof: It is instructive to draw a picture of the consecutive series
of neighborhoods of a fixed node for e.g. a 2-dimensional Bravais
lattice. It is obvious and can also be proved that for m sufficiently
large the number of nodes in $\U_m$ goes like a power of m with the
exponent being the embedding dimension $D_E$ as the euclidean volume
of $\U_m$ grows with the same power.\\[0.5cm]
Remarks:i) For $\U_m$ too small the number of nodes may deviate from
an exact power law which in general becomes only correct for
sufficiently large m.
\\ii) The number of nearest neighbors, on the other side, does not(!)
influence the exponent but rather enters in the prefactor. In other
words, it influences $|\U_m|$ for m small but drops out
asymptotically by taking the logarithm. For an ordinary Bravais
lattice with $N_C$ the number of nodes in a unit cell one has
asymptotically:
\begin{equation} |\U_m|\sim N_C\cdot m^{D_E} \quad\mbox{and hence:}
\end{equation}
\begin{equation} D=\lim_{m\to\infty}(\ln(N_C\cdot m^{D_E})/\ln
m)=D_E+\lim_{m\to\infty}(N_C/\ln m)=D_E 
\end{equation}
independently of $N_C$.\vspace{0.5cm}

Before we proceed a remark should be in order concerning related ideas
on a concept like dimension occurring in however completely different
fields of modern physics:

When we started to work out our own concept we scanned in vain the
literature on e.g. graphs accessible to us and consulted various
experts working in that field. From this we got the impression that
such ideas have not been pursued in that context. (It is however
apparent that there exist conceptual relations to the geometry of {\it
  'fractals'}, whereas in that context '{\it fractal dimension}'
emerges rather by magnifying the microscopic details while in our
situation it is exactly the other way round. These two routes may
however be more closely related if we apply some '{\it renormalisation
  procedure}' to our discrete network in order to reconstruct the
ordinary continuous space-time!).

Quite some time after we developed the above concept we
were kindly informed by Th. Filk that such a concept had been employed
in a however quite different context by e.g. A.A. Migdal et al and by
himself (see e.g. \cite{filk} and \cite{migdal}). Furthermore we found
a cursory remark in \cite{Baxter} on p.49 in connection with the '{\
  Bethe lattice}'. As a consequence one
should say that, while a concept like this may perhaps not be
widely known for discrete structures like ours, it does, on the other
side, not seem to be entirely new. We hope to come back to possible
relations between these various highly interesting approaches
elsewhere (see our remarks before 5.6 Observation). 
 
Matters become much more interesting and subtle if one studies more
general graphs than simple lattices. Note that there exists a general
theorem showing that practically every graph can be embedded in
$\Rl^3$ and still quite a few in $\Rl^2$ ('{\it planar graphs}'). 

The point is however that this embedding is in general not invariant
with respect to the euclidean metric. But something like an apriori
given euclidean length is unphysical for the models we are after anyhow. 
This result has the advantage that one can visualize many graphs
already in, say, $\Rl^2$ whereas their intrinsic dimension may be much
larger.

An extreme example is a '{\it tree graph}', i.e. a graph without
'{\it loops}'. In the following we study an infinite, regular tree
graph with node degree 3, i.e. 3 bonds branching off each node. The
absence of loops means that the '{\it connectivity}' is extremely low
which results in an exceptionally high '{\it dimension}' as we will
see.

Starting from an arbitrary node we can construct the neighborhoods
$\U_m$ and count the number of nodes in $\U_m$ or $\partial\U_m$.
$\U_1$ contains 3 nodes which are linked with the reference node
$n_0$. There are 2 other bonds branching off each of these nodes.
Hence in $\partial\U_2=\U_2\backslash\U_1$ we have $3\cdot2$ nodes
and by induction:
\begin{equation} |\partial\U_{m+1}|=3\cdot2^m \end{equation}
which implies
\begin{equation} D-1:=\lim_{m\to\infty}(\ln|\partial\U_{m+1}|/\ln m)=
\lim_{m\to\infty}(m\cdot ln 2/\ln m+3/\ln m)=\infty \end{equation}
Hence we have: \\[0.5cm]
{\bf 5.7 Observation(Tree)}: The intrinsic dimension of an infinite
tree is $\infty$ and the number of new nodes grows exponentially like
some $n(n-1)^m$  (with $n$ being the node degree).\\[0.5cm]
Remark: $D=\infty$ is mainly a result of the absence of loops(!), in
other words: there is exactly one path, connecting any two nodes. 
This is usually not so in other graphs, e.g. lattices, where the
number of new nodes grows at a much slower pace (whereas the number
of nearest neighbors can nevertheless be large). This is due to the
existence of many loops s.t. many of the nodes which can be reached
from, say, a node of $\partial\U_m$ by one step are already
contained in $\U_m$ itself. \vspace{0.5cm}

We have seen that for, say, lattices the number of new nodes grows
like some fixed power of m while for, say, trees m occurs in the
exponent. The borderline can be found as follows:\\[0.5cm]
{\bf 5.8 Observation}: If for $m\to\infty$ the average number of 
nodes in $\U_{m+1}$ per node contained in $\U_m$ is uniformly away
from zero or, stated differently:
\begin{equation} |\U_{m+1}|/|\U_m|\geq 1+\varepsilon \end{equation}
for some $\varepsilon\geq 0$ then we have exponential growth, in other
words, the intrinsic dimension is $\infty$.
\\The corresponding result holds with $\U_m$ being replaced by
$\partial\U_m$. \\[0.5cm]
Proof: If the above estimate holds for all $m\geq m_0$ we have by
induction:
\begin{equation} |\U_m|\geq |\U_{m_0}|\cdot (1+\varepsilon)^{m-m_0}
\end{equation} 
\vspace{0.5cm}

Potential applications of this concept of intrinsic dimension are
manifold. Our main goal is it to develop a theory which explains how
our classical space-time and what we like to call the '{\it physical vacuum}'
has emerged from a more primordial and discrete background via some
sort of phase transition.

In this context we can also ask in what sense macroscopic(!)
space-time dimension 4
is exceptional, i.e. whether it is merely an accident or whether
there is a cogent reason for it.

As the plan of this paper is mainly to introduce and develop the necessary
conceptual tools and to pave the ground, the bulk of the investigation
in this particular direction shall be presented elsewhere as it is
part of a detailed analysis of the (statitical) dynamics on networks
as introduced above, their possible phase transitions,
selforganisation, emergence of patterns and the like.

In the following we only want to
supply a speculative and very heuristic argument in favor of
(macroscopic) space-dimension
3 which is only designed to show in what direction such an attempt
could be pursued.\\[0.5cm]
Remark: This does not exclude the existence of possible extra (hidden)
internal dimensions of continuous space-time. Quite the contrary,
these would fit very naturally in our network scheme as a description
of the internal structure of certain subclusters of nodes/bonds which
are supposed to constitute the '{\it physical points}'.\vspace{0.5cm}

We emphasized in section 2 that also the bond states, modelling the
strength of local interactions between neighboring nodes, are in our
model theory dynamical variables. In extreme cases these couplings
may completely die out and/or become {\it 'locked in'} between certain
nodes, depending on the kind of model.
It may now happen that in the course of evolution a local island of
'higher order' (or several of them) emerges via a spontaneous
fluctuation in a, on large scales, unordered and erratically
fluctuating network in which couplings between nodes are switched on
and off more or less randomly.

One important effect of the scenario we have in mind (among others) is
that there may occur now a pronounced near order in this island,
accompanied by an increase in correlation length and an effective
screening of the dangerously large {\it 'quantum fluctuations'} on
Planck scale, while the global state outside remains more or less
structureless. We assume that this will be effected by a reduction of
intrinsic dimension within this island which may become substantially lower
than outside, say, finite as compared to (nearly) infinity.

If this '{\it nucleation center}' is both sufficiently large and its
local state '{\it dynamically favorable}' in a sense to be specified
(note that a concept like '{\it entropy}' or something like that
would be of use here) it may start to unfold  and trigger a global phase
transition.

As a result of this phase transition a relatively smooth and stable
submanifold on a certain coarse-grained scale (alluding to the
language of synergetics we would like to call it an '{\it order parameter
manifold}') may come into being which displays certain properties we
would attribute to space-time.

Under these premises we could now ask what is the probability for
such a specific and sufficiently large spontaneous fluctuation? As we
are at the moment talking about heuristics and qualitative behavior
we make the following thumb-rule-like assumptions:\\[0.5cm]
i) In the primordial network '{\it correlation lengths}' are
supposed to be extremely short (more or less nearest neighbor), i.e.
the strengths of the couplings are fluctuating more or less
independently.\\
ii) A large fluctuation of the above type implies in our picture that
a substantial fraction of the couplings in the island passes a certain
threshold (cf. the models of section 2) i.e. become sufficiently weak/dead  and/or cooperative. The probability per
individual bond for this to happen be p. Let L be the diameter of the
nucleation center, $const\cdot L^d$ the number of nodes or bonds in
this island for some d. The probability for such a fluctuation is
then roughly (cf. i)):
\begin{equation} W_d=const\cdot p^{(L^d)} \end{equation}
iii) We know from our experience with phase transitions that there
are favorable dimensions, i.e. the nucleation centers may fade away
if either they themselves are too small or the dimension of the
system is too small. Apart from certain non-generic models $d=3$ is
typically the threshold dimension.\\
iv) On the other side we can compare the relative probabilities for
the occurrence of sufficiently large spontaneous fluctuations for
various d's. One has:
\begin{equation} W_{d+1}/W_d\sim p^{(L^{d+1})}/p^{(L^d)}=p^{L^d(L-1)} 
\end{equation}
Take e.g. $d=3,\,L=10,\,p=1/2$ one gets:
\begin{equation} W_4/W_3\sim 2^{-(9\cdot10^3)} \end{equation}
In other words, provided that this crude estimate has a grain of
truth in it, one may at least get a certain clue that space-dimension
3 is both the threshold dimension and, among the class of in principle
allowed dimensions (i.e. $d\geq3$) the one with the dominant
probability.


\begin{thebibliography}{99}
\bibitem{1}C.J.Isham: "Conceptual and Geometrical Problems in Quantum
Gravity", Lecture presented at the 1991 Schladning Winter School
\bibitem{2}Contributions in Physica 10D(1984), especially the review
by S.Wolfram
\bibitem{3}S.Kauffman: in "Complexity, Entropy and the Physics of
Information, SFI-Studies Vol.VIII p.151, Ed. W.H.Zurek,
Addison-Wesley 1990
\bibitem{4}K.Zuse: in Act.Leop. Vol.37/1 ("Informatik") 1971 p.133,\\
C.F.von Weizs\"acker: l.c. p.509\\
K.Zuse: Int.J.Th.Phys. 21(1982)589
\bibitem{5}R.Feynman: as quoted in D.Finkelstein: Phys.Rev.
184(1969)1261 or:\\ Int.J.Th.Phys. 21(1982)467; in fact most of the
numbers 3/4, 6/7, 12 are devoted to this topic.
\bibitem{6}D.Finkelstein: Int.J.Th.Phys. 28(1989)1081
\bibitem{7}L.Bombelli, J.Lee, D.Meyer, R.Sorkin: Phys.Rev.Lett.
59(1987)521\\
A.P.Balachandran,G.Bimonte,E.Ercolesi,G.Landi,F.Lizzi,G.Sparano,P.Teotonio-Sobrinho:Journ.Geom.Phys.
18(1996)163
\bibitem{8}A.Dimakis, F.M\"uller-Hoissen: J.Math.Phys. 35(1994)6703,\\
F.M\"uller-Hoissen: "Physical Aspects of Differential Calculi on
Commutative Algebras", Karpacz Winter School 1994,\\
H.C.Baehr,A.Dimakis,F.Mueller-Hoissen: Journ.Phys. A 28(1995)3197
\bibitem{9}G.'t Hooft: J.Stat.Phys. 53(1988)323,\\
Nucl.Phys. B342(1990)471
\bibitem{weinberg}S.Weinberg: ''dreams of a final theory'', Vintage,
London 1993
\bibitem{10}M.Requardt: Preprints G\"ottingen resp. in preparation
\bibitem{mack}G.Mack: ''Universal Dynamics of Complex Adaptive Systems
: Gauge Theory of Things Alive'', Desy 94-075 or hep-lat 9411059  
\bibitem{11}O.Ore:"Theory of Graphs", American Math. Soc., N.Y. 1962
\bibitem{12}B.Bollobas:''Graph Theory'', Graduate Texts in
  Mathematics, Springer, N.Y. 1979
\bibitem{schmidt}G.Schmidt, Th.Stroehlein: ''Relationen und Graphen'',
Springer, N.Y. 1989  
\bibitem{toffoli}T.Toffoli, N.Margolus: ''Cellular Automaton
  Machines'', MIT Press, Cambridge Massachusetts 1987
\bibitem{13}A.Connes: "Non-Commutative Geometry", Acad.Pr., N.Y. 1994
\bibitem{14}J.Madore: "Non-Commutative Differential Geometry and its
Physical Applications", LPTHE Orsay 1994
\bibitem{15}P.Seibt: "Cyclic Homology of Algebras", World Scientific,
Singapore 1987
\bibitem{16}D.Kastler: "Cyclic Cohomology within the Differential
Envelope", Hermann, Paris 1988
\bibitem{coque}R.Coqueraux: Journ.Geom.Phys. 6(1989)425 and loc.cit. 11(1993)307
\bibitem{17}S.Lefschetz: "Applications of Algebraic Topology",
Springer, N.Y. 1975
\bibitem{18}G.A.Edgar: "Measure, Topology, and Fractal Geometry",
Springer, N.Y. 1990,\\
K.Kuratowski: "Topology" Vol.1, Acad.Pr., N.Y. 1966
\bibitem{filk}Th.Filk: Mod.Phys.Lett. A7(1992)2637
\bibitem{migdal}M.E.Agishtein,A.A.Migdal: Nucl.Phys. B350(1991)690
\bibitem{Gibbs}P.Gibbs: ''The small scale structure of space-time: a
  bibliographical review'', hep-th/9506171
\bibitem{Planck}M.Requardt: ''Emergence of Space-Time on the
  Planck-Scale described as an Unfolding Phase Transition within the
  Scheme of Dynamical Cellular Networks and Random Graphs,
  hep-th/9610055
\bibitem{Antonsen}F.Antonsen:Int.Journ.Theor.Phys. 33(1994)1189 
\bibitem{Cuntz}A.Connes,J.Cuntz: Commun.Math.Phys. 114(1988)515
\bibitem{Dubois}M.Dubois-Violette,J.Madore,T.Masson,J.Mourad:Journ.Math.Phys.
37(1996)4089
\bibitem{groupoid}Groupoid Home Page:
http://amath-www.Colorado.edu:80/math/research
groups/groupoids/groupoids.shtml, 
e.g. the paper by A.Weinstein:''Groupoids: Unifying Internal and
External Symmetry''.\\
Another source is the textebook by R.Brown\\
R.Brown:''Topology: a geometric account of general topology\ldots'',
Halsted Pr., N,Y. 1988\\
or the Lecture Note by J.Renault\\
J.Renault:''A Groupoid Approach to C*-algebras'', Lect.Notes in
Math. 793, Springer, Heidelberg 1980
\bibitem{Baxter}R.J.Baxter:''Exactly Solvable Models in Statistical
  Mechanics'', Acad.Pr., N.Y. 1982



\end{thebibliography}
\end{document}